\documentclass[12pt,a4paper]{article}
\usepackage{a4wide,cite}
\usepackage{amsmath,amssymb}
\usepackage{latexsym,graphicx,epsfig}

\newcommand{\half}{{\textstyle\frac{1}{2}}}

\newcommand{\fourth}{{\textstyle\frac{1}{4}}}
\newcommand{\sixth}{{\textstyle\frac{1}{6}}}

\def\lsim{\mathrel{\rlap{\raise 2.5pt \hbox{$<$}}\lower 2.5pt\hbox{$\sim$}}}
\def\gsim{\mathrel{\rlap{\raise 2.5pt \hbox{$>$}}\lower 2.5pt\hbox{$\sim$}}}

\renewcommand{\Re}{{\rm Re\thinspace}}
\renewcommand{\Im}{{\rm Im\thinspace}}

\allowdisplaybreaks %!!!!
\begin{document}
\renewcommand{\thefootnote}{\fnsymbol{footnote}}
\newpage\normalsize
    \pagestyle{plain}
    \setlength{\baselineskip}{4ex}\par
    \setcounter{footnote}{0}
    \renewcommand{\thefootnote}{\arabic{footnote}}
\newcommand{\preprint}[1]{%
\begin{flushright}
\setlength{\baselineskip}{3ex} #1
\end{flushright}}
\renewcommand{\title}[1]{%
\begin{center}
    \LARGE #1
\end{center}\par}
\renewcommand{\author}[1]{%
\vspace{2ex}
{\Large
\begin{center}
    \setlength{\baselineskip}{3ex} #1 \par
\end{center}}}
\renewcommand{\thanks}[1]{\footnote{#1}}
\begin{flushright}
    \today
\end{flushright}
\vskip 0.5cm

\begin{center}
{\bf \Large
Maximal CP nonconservation \\
in the Two-Higgs-Doublet Model\footnote{Invited talk given at the 
      Cracow Epiphany conference on heavy flavours,
      3 - 6 January 2003, Cracow, Poland.
      To be published in Acta Physica Polonica, July 2003.}}
\end{center}
\vspace{1cm}
\begin{center}
Wafaa Khater  {\rm and}
Per Osland
\end{center}
%-----------------------------------
%   Address
%-----------------------------------
\vspace{1cm}
\begin{center}
Department of Physics,
University of Bergen,
    Allegt.\ 55, N-5007 Bergen, Norway
\end{center}
\vspace{1cm}
%%%%%%%%%%%%%%%%%%%%%%%%%%%%%%%%%%%%%%%%%%%%%%%%
\begin{abstract}
We study the simplest Two-Higgs-Doublet Model
that allows for CP nonconservation, where it can be parametrized by only {\it
one} parameter in the Higgs potential.  Different concepts of {\it maximal}
CP-nonconservation in the gauge-Higgs and the quark-Higgs (Yukawa) sectors are
compared.  Maximal CP nonconservation in the gauge-Higgs sector does normally
not lead to maximal CP nonconservation in the Yukawa sector, and vice versa.
\end{abstract}
%%%%%%%%%%%%%%%%%%%%%%%%%%%%%%%%%%%%%%%%%%%%%%%%%%%%%%%%%%%%%%%%%%%%%%%%
\section{Introduction}
\setcounter{equation}{0}
%%%%%%%%%%%%%%%%%%%%%%%%%%%%%%%%%%%%%%%%%%%%%%%%%%%%%%%%%%%%%%%%%%%%%%%%
Mendez and Pomarol introduced the concept of maximal CP nonconservation
\cite{Mendez:1991gp} in the context of the gauge--Higgs sector of the
Two-Higgs-Doublet Model (2HDM) \cite{Lee:iz}.  In the absence of CP
nonconservation, only two of the three neutral Higgs bosons couple to the
electroweak gauge bosons (the two CP even ones, often denoted $h$ and $H$).
When CP is {\it not} conserved, all three do. In fact, Mendez and Pomarol
realized that the product of all three gauge--Higgs couplings, which is
bounded by unitarity, is a useful concept to parametrize the amount of CP
nonconservation, and defined the quantity
\begin{equation} \label{Eq:xi-v-def}
\xi_V=27[g_{VVH_1}\,g_{VVH_2}\,g_{VVH_3}]^2
\end{equation}
as a measure of CP nonconservation in the gauge--Higgs sector.
If the couplings $g_{VVH_i}$ are normalized with respect to those
of the Standard Model (SM), then $\xi_V$, as defined above, satisfies
\begin{equation}
0\le \xi_V \le 1.
\end{equation}
However, this measure of CP nonconservation is not applicable
to the fermion--Higgs sector.

In the fermion--Higgs sector of a given version of the 2HDM, one should
consider quantities other than $\xi_V$ as measures of CP nonconservation. As
we will see from our investigation, the parameters of the 2HDM that maximize
$\xi_V$ are {\it different} from those that maximize CP nonconservation in
the Yukawa sector.  They are in general also different for
the up- and down-quark sectors.

The paper is organized as follows.
In sect.~2 we review the 2HDM and in sect.~3 we study the conditions for
maximum CP nonconservation in the gauge--Higgs sector.
Sections ~4 and 5 are devoted to the Yukawa sector,
at the parton and proton level, respectively, and sect.~6
contains some concluding remarks.
%%%%%%%%%%%%%%%%%%%%%%%%%%%%%%%%%%%%%%%%%%%%%%%%%%%%%%%%%%%%%%%%%%%%%%%%
\section{The Two-Higgs-Doublet Model}  \label{sec:2HDM}
\setcounter{equation}{0}
%%%%%%%%%%%%%%%%%%%%%%%%%%%%%%%%%%%%%%%%%%%%%%%%%%%%%%%%%%%%%%%%%%%%%%%%
We shall here introduce some notation for the Two-Higgs-Doublet Model
\cite{HHG}.
Let the Higgs potential be parametrized as \cite{Ginzburg:2001ss}
\begin{eqnarray}                    \label{Eq:gko-pot}
V&=&\frac{\lambda_1}{2}(\phi_1^\dagger\phi_1)^2
+\frac{\lambda_2}{2}(\phi_2^\dagger\phi_2)^2
+\lambda_3(\phi_1^\dagger\phi_1) (\phi_2^\dagger\phi_2)
+\lambda_4(\phi_1^\dagger\phi_2) (\phi_2^\dagger\phi_1) \\
&&+\frac{1}{2}\left[\lambda_5(\phi_1^\dagger\phi_2)^2+{\rm h.c.}\right]
-\frac{1}{2}\left\{m_{11}^2(\phi_1^\dagger\phi_1)
+\left[m_{12}^2 (\phi_1^\dagger\phi_2)+{\rm h.c.}\right]
+m_{22}^2(\phi_2^\dagger\phi_2)\right\}. \nonumber 
\end{eqnarray}
The parameters $\lambda_5$ and $m_{12}^2$ are allowed to be complex,
subject to the constraint
\begin{equation}
\Im m_{12}^2=\Im \lambda_5\, v_1 v_2,
\end{equation}
with $v_1$ and $v_2$ the vacuum expectation values ($v_1^2+v_2^2=v^2$,
with $v=246\text{ GeV}$).

The corresponding neutral-Higgs mass matrix squared is then
given by
\begin{equation}          \label{Eq:MM}
{\cal M}=v^2
\begin{pmatrix}
\lambda_1c_\beta^2+\nu s_\beta^2&
(\lambda_{345} -\nu)c_\beta s_\beta&
-\half\Im\lambda_5\, s_\beta\\[4mm] 
(\lambda_{345} -\nu)c_\beta s_\beta&
\lambda_2s_\beta^2+\nu c_\beta^2&
-\half\Im\lambda_5\, c_\beta\\[4mm] 
-\half\Im\lambda_5\, s_\beta& 
-\half\Im\lambda_5\, c_\beta&
-\Re\,\lambda_5+\nu
\end{pmatrix}
\end{equation}
with the abbreviations $c_\beta=\cos\beta$, $s_\beta=\sin\beta$,
$\tan\beta=v_2/v_1$, $\lambda_{345}=\lambda_3+\lambda_4+\Re\lambda_5$,
$\nu=\Re m_{12}^2/(2v^2\sin\beta\cos\beta)$ and $\mu^2=v^2\nu$.

The $(1,3)$ and $(2,3)$ elements of this mass-squared matrix (\ref{Eq:MM}),
which are responsible for CP nonconservation, are related
via the angle $\beta$. In this sense, CP nonconservation is described
by {\it one} parameter, namely $\Im\lambda_5$.

In order to diagonalize this matrix (\ref{Eq:MM}), we introduce the rotation
matrix
\begin{align}     \label{Eq:R-angles}
R=R_c\,R_b\,R_{\tilde\alpha}
=&\begin{pmatrix}
1         &    0         &    0 \\
0 &  \cos\alpha_c & \sin\alpha_c \\
0 & -\sin\alpha_c & \cos\alpha_c
\end{pmatrix}
\begin{pmatrix}
\cos\alpha_b & 0 & \sin\alpha_b \\
0         &       1         & 0 \\
-\sin\alpha_b & 0 & \cos\alpha_b
\end{pmatrix}
\begin{pmatrix}
\cos\tilde\alpha & \sin\tilde\alpha & 0 \\
-\sin\tilde\alpha & \cos\tilde\alpha & 0 \\
0         &       0         & 1
\end{pmatrix}  \nonumber \\
=&\begin{pmatrix}
c_{\tilde\alpha}\,c_b & s_{\tilde\alpha}\,c_b & s_b \\
- (c_{\tilde\alpha}\,s_b\,s_c + s_{\tilde\alpha}\,c_c) 
& c_{\tilde\alpha}\,c_c - s_{\tilde\alpha}\,s_b\,s_c & c_b\,s_c \\
- c_{\tilde\alpha}\,s_b\,c_c + s_{\tilde\alpha}\,s_c 
& - (c_{\tilde\alpha}\,s_c + s_{\tilde\alpha}\,s_b\,c_c) & c_b\,c_c
\end{pmatrix}
\end{align}
with $c_i=\cos\alpha_i$, $s_i=\sin\alpha_i$, and satisfying
\begin{equation} \label{Eq:diagonal}
R{\cal M}R^{\rm T}={\rm diag}(M_1^2,M_2^2,M_3^2).
\end{equation}
Here, $M_1\le M_2\le M_3$.
The angular ranges are taken as
$-\pi/2<\tilde\alpha\le\pi/2$, $-\pi<\alpha_b\le\pi$, and
$-\pi/2<\alpha_c\le\pi/2$.
As discussed in \cite{Khater:2003wq}, only some regions of the parameter
space are physically allowed.

%%%%%%%%%%%%%%%%%%%%%%%%%%%%%%%%%%%%%%%%%%%%%%%%%%%%%%%%%%%%%%%%%%%%%%%%
\begin{figure}[htb]
\refstepcounter{figure}
\label{Fig:albc-100-500-600-300}
\addtocounter{figure}{-1}
\begin{center}
\setlength{\unitlength}{1cm}
\begin{picture}(15.0,12.0)
\put(0,-0.7)
{\mbox{\epsfysize=13cm
 \epsffile{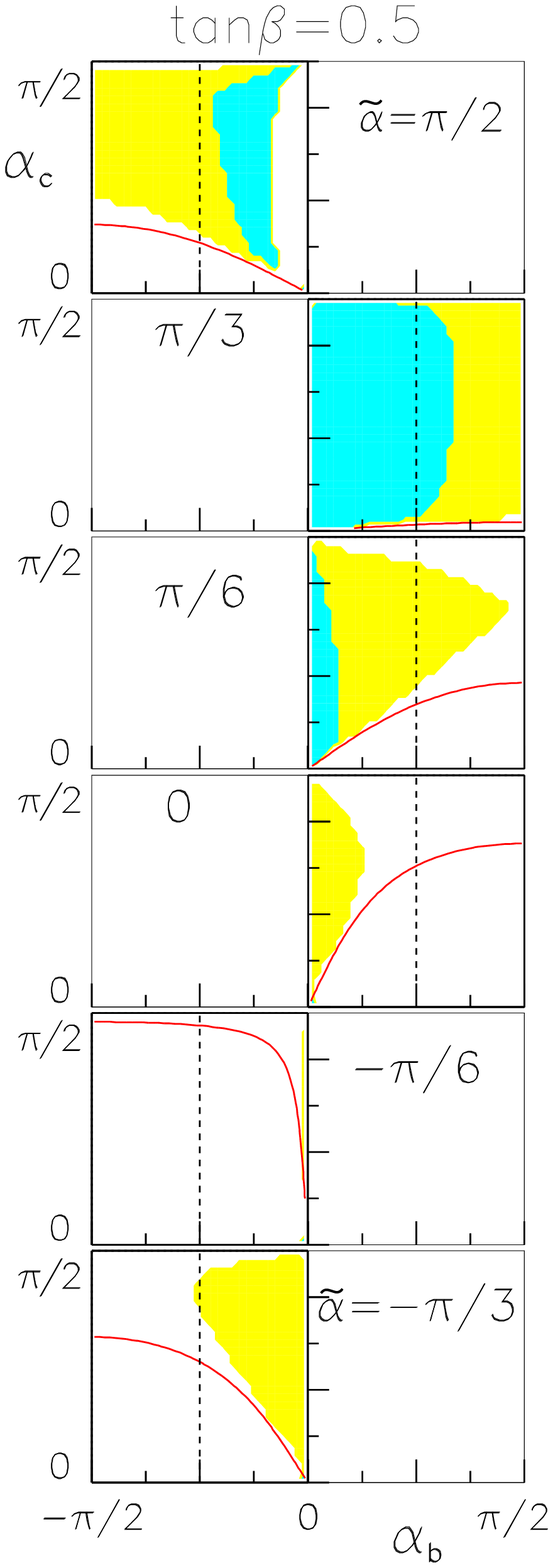}}
 \mbox{\epsfysize=13cm
 \epsffile{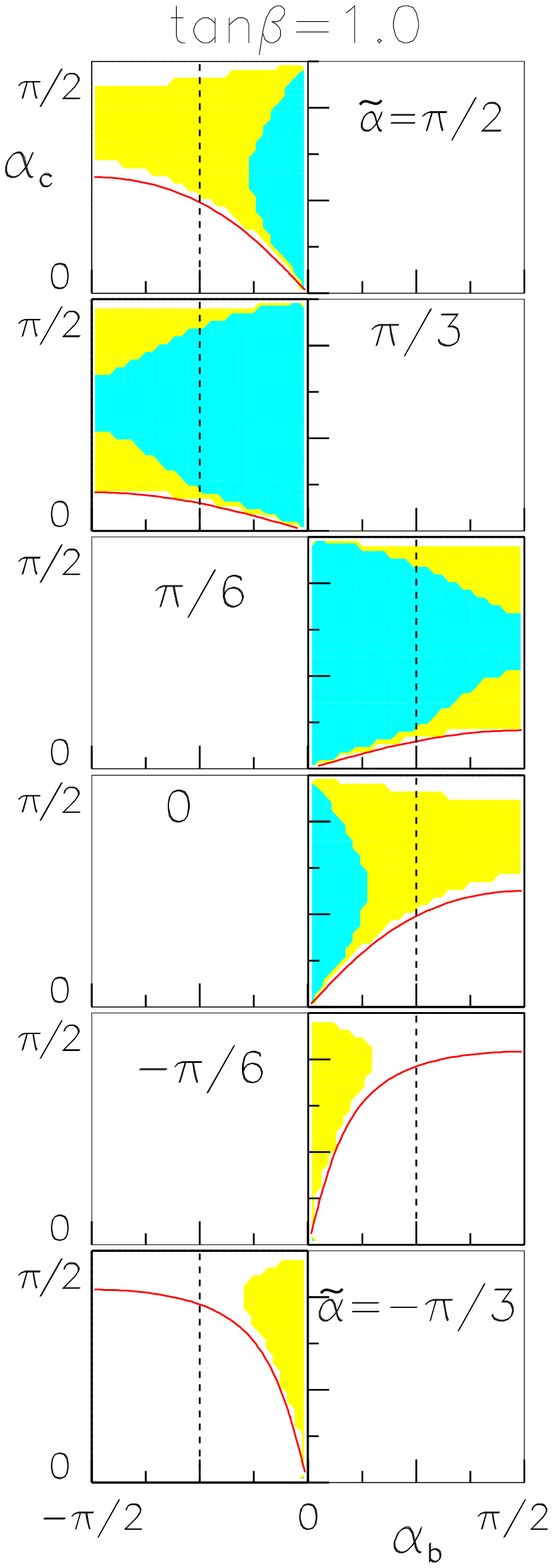}}
 \mbox{\epsfysize=13cm
 \epsffile{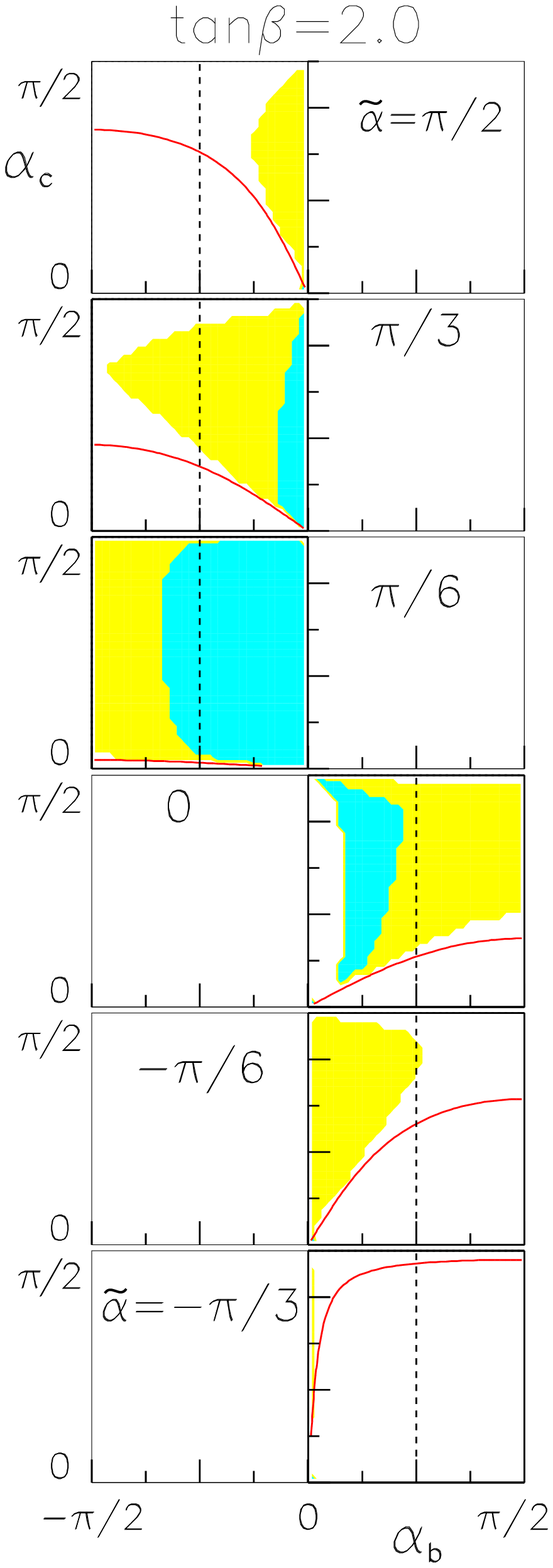}}}
\end{picture}
%\vspace*{-4mm}
\caption{Dark (blue): physical regions [see Eq.~(\ref{Eq:xi-pert})] in the
$\alpha_b$--$\alpha_c$ plane for various values of $\tan\beta$ and
$\tilde\alpha$.  Soft parameters: $M_1=100~\text{GeV}$, $M_2=500~\text{GeV}$,
$M_{H^\pm}=600~\text{GeV}$, $\mu=300~\text{GeV}$, $\xi_{\rm pert}=1$.  Light
(yellow): Same with $\xi_{\rm pert}=5$. Solid contour: absolute boundary.}
\end{center}
\end{figure}
%%%%%%%%%%%%%%%%%%%%%%%%%%%%%%%%%%%%%%%%%%%%%%%%%%%%%%%%%%%%%%%%%%%%%%

This limitation of the parameter space is due to various constraints,
including (i) $M_1\le M_2\le M_3$, and (ii) the constraints of
perturbativity and unitarity. We shall represent the latter as 
\begin{equation} \label{Eq:xi-pert}
|\lambda_i|< 4\pi\,\xi_{\rm pert},\quad \text{with }
\xi_{\rm pert} = {\cal O}(1).
\end{equation}

We show in Fig.~\ref{Fig:albc-100-500-600-300} typical allowed regions in the
$\alpha_b$--$\alpha_c$ plane, for a few values of $\tan\beta$ and
$\tilde\alpha$. In this figure, we only show regions of $|\alpha_b|\le\pi/2$
and only positive $\alpha_c$.  Regions of larger $|\alpha_b|$ and negative
$\alpha_c$ are given by the symmetries discussed in
\cite{Khater:2003wq}. Furthermore, for given values of $\tan\beta$ and
$\tilde\alpha$ (and given sign of $\alpha_c>0$), only {\it one sign} of
$\alpha_b$ is realized, requiring $M_2\le M_3$.  The dashed lines at
$\alpha_b=\pm\pi/4$ indicate where CP nonconservation is maximal in the
Higgs--top-quark sector, in the limit of {\it one} light Higgs boson and two
heavier ones, see Eq.~(\ref{Eq:gamma_CP-1}).

Different choices for the `soft parameters' (in particular, different values
of $\mu^2$) lead to somewhat different allowed regions. Also, a larger value
of $\xi_{\rm pert}$ extends the region. However, there are absolute bounds,
indicated by the solid contours outside the shaded regions in
Fig.~\ref{Fig:albc-100-500-600-300}, that can not be crossed for any choice of
the `soft parameters' \cite{Khater:2003wq}. In order to cover a range of
different choices for $\mu^2$, one may take a rather large value of $\xi_{\rm
pert}$ (in sect.~5 we shall consider $\xi_{\rm pert}=5$).  For further
discussion of these issues, see \cite{Khater:2003wq,GKO}.

In this notation, Eqs.~(\ref{Eq:MM})--(\ref{Eq:diagonal}), the gauge--Higgs
couplings are, relative to the corresponding SM coupling,  given by
\begin{equation} \label{Eq:HZZ-couplings}
H_i ZZ: \qquad g_{VVH_i}=\cos\beta\, R_{i1} +\sin\beta\, R_{i2},
\end{equation}
whereas for the Yukawa couplings we consider the so-called Model II \cite{HHG}
where they are given by
\begin{align}  \label{Eq:H_itt}
H_j t \bar t: \qquad 
&\frac{1}{\sin\beta}\, [R_{j2}-i\gamma_5\cos\beta R_{j3}]
\equiv[a_j^{(t)}+i\gamma_5\tilde a_j^{(t)}], \\
H_j b \bar b: \qquad 
&\frac{1}{\cos\beta}\, [R_{j1}-i\gamma_5\sin\beta R_{j3}]
\equiv[a_j^{(b)}+i\gamma_5\tilde a_j^{(b)}],
\label{Eq:H_ibb}
\end{align}
with $R_{ij}$ an element of the rotation matrix (\ref{Eq:R-angles}).
%%%%%%%%%%%%%%%%%%%%%%%%%%%%%%%%%%%%%%%%%%%%%%%%%%%%%%%%%%%%%%%%%%%%%%%%
\section{\boldmath CP nonconservation in the gauge-Higgs sector}
\setcounter{equation}{0}
%%%%%%%%%%%%%%%%%%%%%%%%%%%%%%%%%%%%%%%%%%%%%%%%%%%%%%%%%%%%%%%%%%%%%%%%

In the gauge--Higgs sector, the amount of CP nonconservation [cf.\
Eq.~(\ref{Eq:xi-v-def})] is in the above notation given by
\begin{equation}  \label{Eq:xi-V-R}
\xi_V=27\prod_{i=1}^3[\cos\beta R_{i1}+\sin\beta R_{i2}]^2.
\end{equation}
This $\xi_V$ depends on $\tan\beta$ as well as on the three angles
$\tilde\alpha$, $\alpha_b$ and $\alpha_c$ that determine $R_{ij}$.  However,
it only depends on $\beta$ and $\tilde\alpha$ through their {\it difference}.
In fact, using (\ref{Eq:R-angles}) and some trigonometric identities, we find
\begin{equation}   \label{Eq:xi_param}
\xi_V
=27c_b^2\cos^2(\beta-\tilde\alpha)
      [s_b\,s_c\cos(\beta-\tilde\alpha) -c_c\sin(\beta-\tilde\alpha)]^2
      [s_b\,c_c\cos(\beta-\tilde\alpha) +s_c\sin(\beta-\tilde\alpha)]^2.
\end{equation}
It is also seen that $\xi_V$
is unchanged under
\begin{alignat}{2} \label{Eq:xi-symm}
&\alpha_b~\text{fixed},\quad
(\alpha_c\leftrightarrow\pi/2+\alpha_c): &\qquad 
&\xi_V\leftrightarrow\xi_V, \nonumber \\
(&\alpha_b\leftrightarrow-\alpha_b),\quad
(\alpha_c\leftrightarrow\pi/2-\alpha_c): &\qquad 
&\xi_V\leftrightarrow\xi_V.
\end{alignat}

In order to provide some intuition for how the CP nonconservation depends on
the parameters of the 2HDM, we show in Fig.~\ref{Fig:xi-V-albc} contours of
constant $\xi_V$ in the $\alpha_b$--$\alpha_c$ plane, for various values of
$\tan\beta$ and $\tilde\alpha$.  We note that there is little CP
nonconservation for `large' values of $\alpha_b$, because of the factor
$c_b^2$ in (\ref{Eq:xi_param}).  Also, there is CP nonconservation even for
$\alpha_b=0$ and for $\alpha_c=0$ (but not when both vanish).
%%%%%%%%%%%%%%%%%%%%%%%%%%%%%%%%%%%%%%%%%%%%%%%%%%%%%%%%%%%%%%%%%%%%%%%%
\begin{figure}[htb]
\refstepcounter{figure}
\label{Fig:xi-V-albc}
\addtocounter{figure}{-1}
\begin{center}
\setlength{\unitlength}{1cm}
\begin{picture}(15.0,12.0)
\put(0,-0.7)
{\mbox{\epsfysize=13cm
 \epsffile{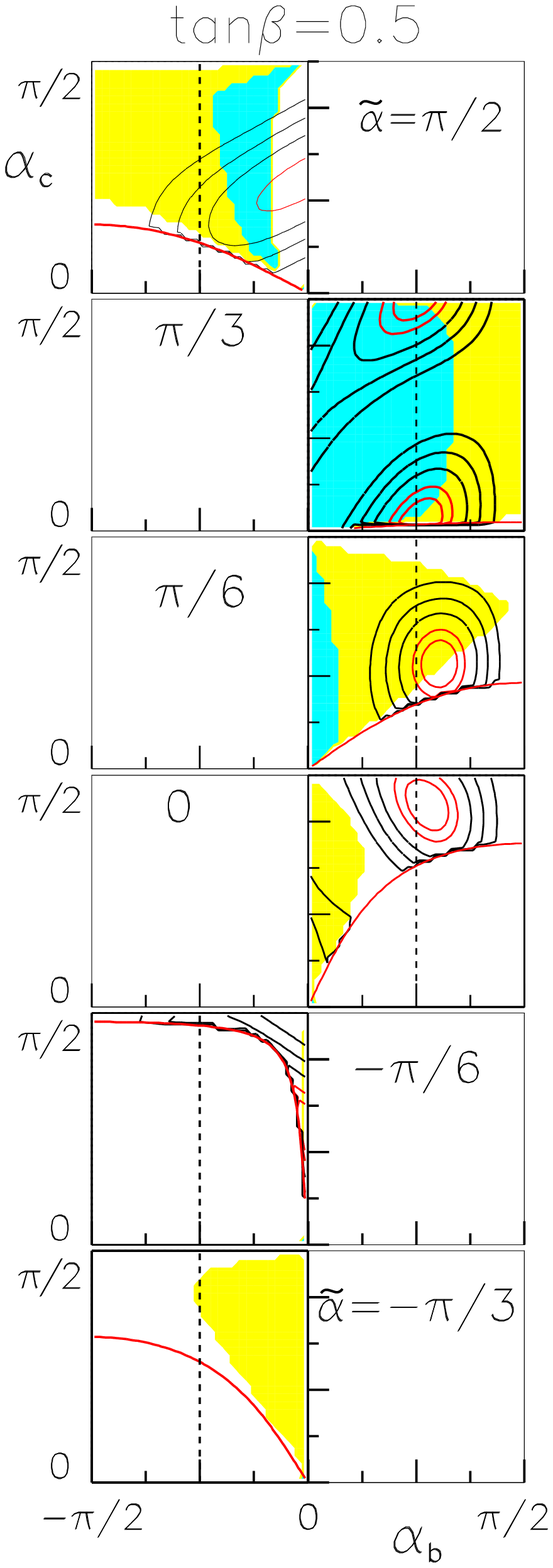}}
 \mbox{\epsfysize=13cm
 \epsffile{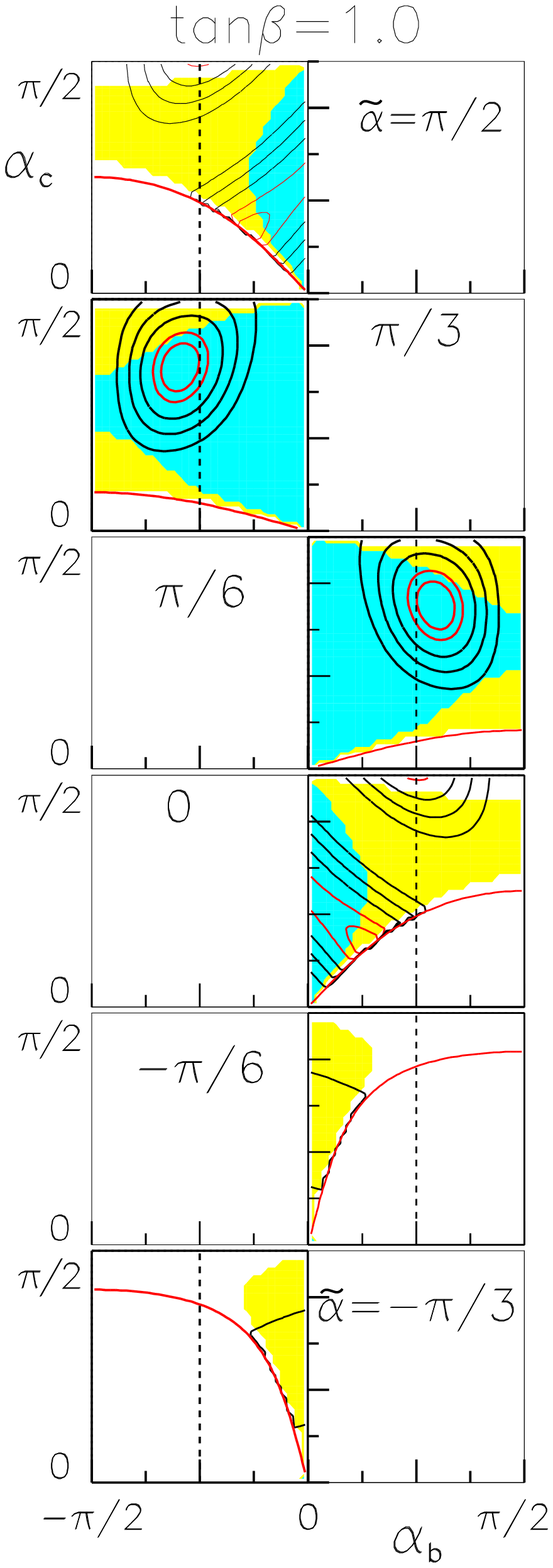}}
 \mbox{\epsfysize=13cm
 \epsffile{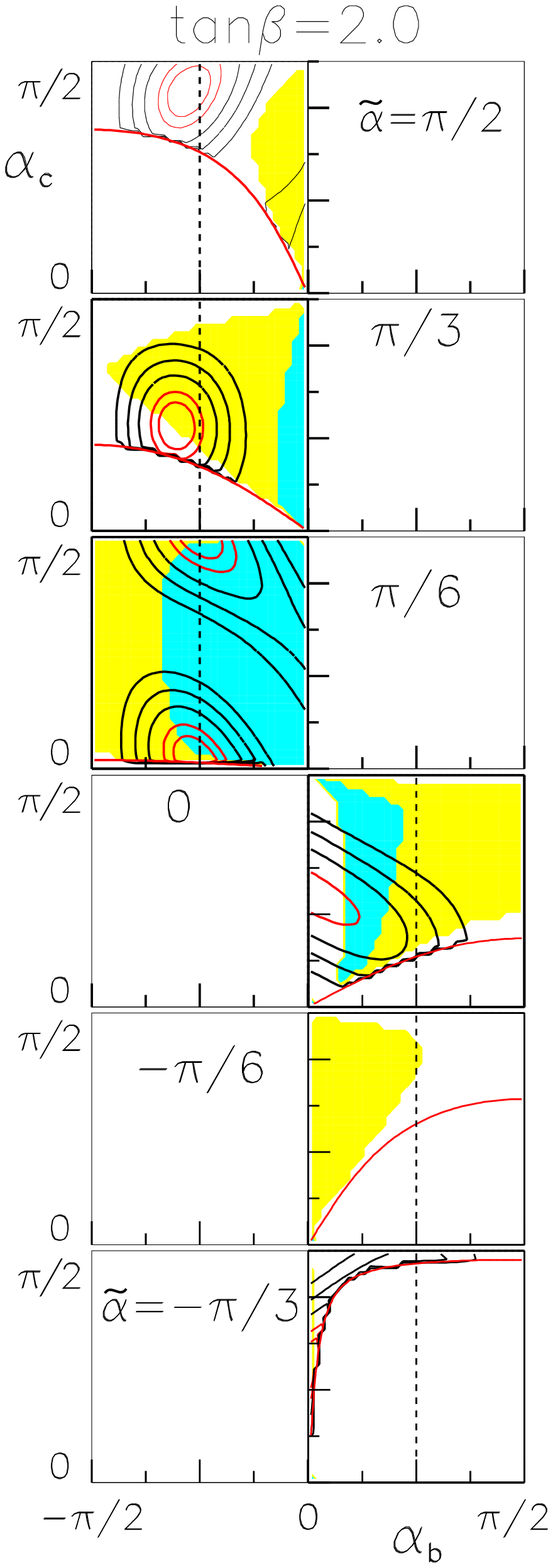}}}
\end{picture}
%\vspace*{-4mm}
\caption{Contours of constant $\xi_V$ [see Eq.~(\ref{Eq:xi-V-R})] in the
$\alpha_b$--$\alpha_c$ plane for various values of $\tan\beta$ and
$\tilde\alpha$.  Soft parameters: $M_1=100~\text{GeV}$, $M_2=500~\text{GeV}$,
$M_{H^\pm}=600~\text{GeV}$, $\mu=300~\text{GeV}$. Dark (blue): $\xi_{\rm
pert}=1$, light (yellow): $\xi_{\rm pert}=5$.}
\end{center}
\end{figure}
%%%%%%%%%%%%%%%%%%%%%%%%%%%%%%%%%%%%%%%%%%%%%%%%%%%%%%%%%%%%%%%%%%%%%%

%%%%%%%%%%%%%%%%%%%%%%%%%%%%%%%%%%%%%%%%%%%%%%%%%%%%%%%%%%%%%%%%%%%%%%%%
\subsection{Simple limits}
%%%%%%%%%%%%%%%%%%%%%%%%%%%%%%%%%%%%%%%%%%%%%%%%%%%%%%%%%%%%%%%%%%%%%%%%
It is instructive to consider the simple limits of $\alpha_b=0$
or $\alpha_c=0$.

\subsubsection*{\boldmath $\alpha_b=0$}

For $\alpha_b=0$, the rotation matrix simplifies:
\begin{equation}     \label{Eq:R-sb=0}
R=\begin{pmatrix}
c_{\tilde\alpha} & s_{\tilde\alpha} & 0 \\
- s_{\tilde\alpha}\,c_c & c_{\tilde\alpha}\,c_c  & s_c \\
  s_{\tilde\alpha}\,s_c & - c_{\tilde\alpha}\,s_c & c_c
\end{pmatrix},
\end{equation}
and one finds
\begin{equation}
\xi_V(\alpha_b=0)=\frac{27}{4}\,
\sin^2(2\alpha_c)\sin^4(\beta-\tilde\alpha)\cos^2(\beta-\tilde\alpha).
\end{equation}
The maximum is given by
\begin{equation} \label{Eq:max-alb=0}
\xi_V=1\quad\text{for}\quad
\tilde\alpha=\beta\pm\arctan\sqrt{2},\quad
\alpha_b=0, \quad
\alpha_c=\pm\pi/4.
\end{equation}

\subsubsection*{\boldmath $\alpha_c=0$}

For $\alpha_c=0$, one finds
\begin{equation}
\xi_V(\alpha_c=0)=\frac{27}{4}\,
\sin^2(2\alpha_b)\cos^4(\beta-\tilde\alpha)\sin^2(\beta-\tilde\alpha).
\end{equation}
This relation holds also for $\alpha_c=\pi/2$.
The maximum is given by
\begin{equation} \label{Eq:max-alc=0}
\xi_V=1\quad\text{for}\quad
\tilde\alpha=\beta\pm\arctan(1/\sqrt{2}),\quad
\alpha_b=\pm\pi/4, \quad
\alpha_c=0\ \text{or}\ \alpha_c=\pi/2.
\end{equation}
%%%%%%%%%%%%%%%%%%%%%%%%%%%%%%%%%%%%%%%%%%%%%%%%%%%%%%%%%%%%%%%%%%%%%%%%
\subsection{\boldmath Maxima of $\xi_V$}
%%%%%%%%%%%%%%%%%%%%%%%%%%%%%%%%%%%%%%%%%%%%%%%%%%%%%%%%%%%%%%%%%%%%%%%%

Since maximizing over angles allows us to keep {\it two} Higgs masses fixed
\cite{Khater:2003wq} and since by Eq.~(\ref{Eq:xi_param}), the
dependence of $\xi_V$\, on $\beta$ and $\tilde\alpha$ shows up in the
form ($\beta-\tilde\alpha$), $\xi_V$ can be maximized for fixed
($\beta-\tilde\alpha$) by meeting the two conditions: 
\begin{equation} \label{Eq:xi_V_bc=0}
\frac{\partial\xi_V}{\partial\alpha_b}=0 \qquad \text{and} \qquad 
\frac{\partial\xi_V}{\partial\alpha_c}=0.
\end{equation}
By substituting from Eq.~(\ref{Eq:xi_param}), and solving
(\ref{Eq:xi_V_bc=0}) for $\alpha_b$ and $\alpha_c$, we obtain a
continuum of maxima:  
\begin{align}  \label{Eq:maximum-xi-v}
\xi_V=1\quad\text{for}\quad
\alpha_b&=\pm\arccos\sqrt{\frac{1+\tan^2(\beta-\tilde\alpha)}{3}}, 
\nonumber \\
\alpha_c&=\pm\arctan
\frac{1+\tan^2(\beta-\tilde\alpha)
-\sqrt{3[2-\tan^2(\beta-\tilde\alpha)]}\,\tan(\beta-\tilde\alpha)}
             {2\tan^2(\beta-\tilde\alpha)-1}.
\end{align}
which impose the constraint 
\begin{equation}  \label{Eq:max-range}
|\tan(\beta-\tilde\alpha)|\le\sqrt{2}
\end{equation}
on ($\beta-\tilde\alpha$).
We note that (\ref{Eq:max-alb=0}) and (\ref{Eq:max-alc=0}) are both special
cases of this (\ref{Eq:maximum-xi-v}).

We show in Fig.~\ref{Fig:angles-ximax} how these angles
$\alpha_b$ and $\alpha_c$ vary with $\tan\beta$ (for fixed $\tilde\alpha$)
when we maximize $\xi_V$.
For a given value of $\tilde\alpha$, these curves only cover a finite range
in $\tan\beta$. They are cut off by (\ref{Eq:max-range}), which
says that, in order to have $\xi_V=1$, $\beta$ and $\tilde\alpha$ should not
differ by more than $\arctan\sqrt2\simeq 54.7\,{}^\circ$. 
In addition, they are cut off by the condition
of having a physical solution as discussed in sect.~\ref{sec:2HDM}, and
delineated by the solid contours in Fig.~\ref{Fig:albc-100-500-600-300}.
Note that there are also solutions having other signs for $\alpha_b$
and $\alpha_c$, but that the model is only physically consistent
for certain sign combinations.
%%%%%%%%%%%%%%%%%%%%%%%%%%%%%%%%%%%%%%%%%%%%%%%%%%%%%%%%%%%%%%%%%%%%%%%%
\begin{figure}[htb]
\refstepcounter{figure}
\label{Fig:angles-ximax}
\addtocounter{figure}{-1}
\begin{center}
\setlength{\unitlength}{1cm}
\begin{picture}(10.0,6.5)
\put(0,0){
\mbox{\epsfysize=7cm
\epsffile{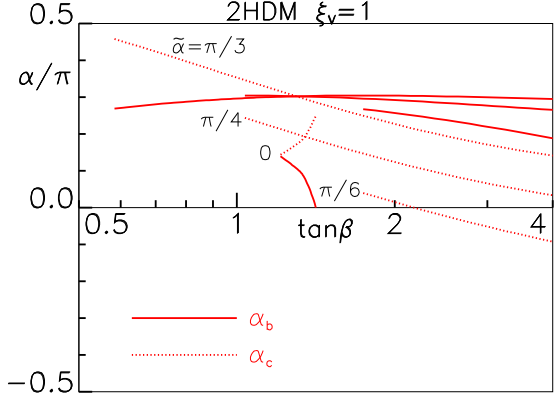}}}
\end{picture}
\vspace*{-4mm}
\caption{Angles $\alpha_b$ and $\alpha_c$ [cf.\ Eq.~(\ref{Eq:maximum-xi-v})]
for which the CP nonconservation $\xi_V$ in the gauge-Higgs sector is maximal,
for a range of $\tan\beta$ values, and for $\tilde\alpha=0$, $\pi/6$, $\pi/4$,
and $\pi/3$.}
\end{center}
\end{figure}
%%%%%%%%%%%%%%%%%%%%%%%%%%%%%%%%%%%%%%%%%%%%%%%%%%%%%%%%%%%%%%%%%%%%%%

%%%%%%%%%%%%%%%%%%%%%%%%%%%%%%%%%%%%%%%%%%%%%%%%%%%%%%%%%%%%%%%%%%%%%%%%
\section{\boldmath CP nonconservation in the Yukawa sector}
\setcounter{equation}{0}
%%%%%%%%%%%%%%%%%%%%%%%%%%%%%%%%%%%%%%%%%%%%%%%%%%%%%%%%%%%%%%%%%%%%%%%%
In the Yukawa sector, one can define measures of CP nonconservation analogous
to the one for the gauge-Higgs sector [cf.\ $\xi_V$ of
Eq.~(\ref{Eq:xi-v-def})].  Requiring thus that {\it all three} Higgs bosons
should have CP-nonconserving couplings to up- and down-type quarks, it is
natural to consider the quantities [see Eqs.~(\ref{Eq:H_itt}),
(\ref{Eq:H_ibb}) and (\ref{Eq:gamma_CP-i})]:
\begin{align}  \label{Eq:xi_t-def}
\xi_t&=\left(\frac{\cos\beta}{\sin^2\beta}\right)^6
\prod_{i=1}^3[R_{i2}\,R_{i3}]^2
\equiv\left(\frac{\cos\beta}{\sin^2\beta}\right)^6\tilde\gamma_t, \nonumber \\
\xi_b&=\left(\frac{\sin\beta}{\cos^2\beta}\right)^6
\prod_{i=1}^3[R_{i1}\,R_{i3}]^2
\equiv\left(\frac{\sin\beta}{\cos^2\beta}\right)^6\tilde\gamma_b.
\end{align}
Both of these differ from the $\xi_V$ defined above in two respects.
First of all, the dependence on $\beta$ factorizes.
Secondly, they individually diverge as $\sin\beta\to0$
(for up-type quarks)
or $\cos\beta\to0$ (for down-type quarks).

One could also consider the quantities
\begin{equation}  \label{Eq:zeta-def}
\zeta_t=\left(\frac{\cos\beta}{\sin^2\beta}\right)^2
\sum_{i=1}^3[R_{i2}\,R_{i3}]^2
\end{equation}
and similarly $\zeta_b$ as measures of CP nonconservation in the Yukawa
sector. These measures---unlike those in (\ref{Eq:xi_t-def})---are consistent
with the fact that if $H_1$ conserves CP in its couplings to the up- and
down-type quarks, i.e.\ $\alpha_b=0$, then the Yukawa sector {\it may still be
CP nonconserving}, since the other two Higgs states, $H_2$ and $H_3$, may have
CP nonconserving couplings to the quarks. Accordingly, $\zeta_t\neq 0$ and
$\zeta_b\neq 0$ for $\alpha_b=0$ which is not the case for $\xi_t$ and
$\xi_b$. This $\zeta_t$ will be discussed in sect.~\ref{subsect:zeta}.

Substituting now from (\ref{Eq:R-angles}) into (\ref{Eq:xi_t-def}), we obtain
for this case of Model~II Yukawa couplings:
\begin{align} \label{Eq:Yukawa-gamma}
\tilde\gamma_t
&= c_b^{6}(s_{\tilde\alpha}\,s_b\,c_c\,s_c)^2
  [s_{\tilde\alpha}\,s_b\,c_c + c_{\tilde\alpha}\,s_c]^2
  [s_{\tilde\alpha}\,s_b\,s_c - c_{\tilde\alpha}\,c_c]^2, \nonumber \\
\tilde\gamma_b
&= c_b^{6}(c_{\tilde\alpha}\,s_b\,c_c\,s_c)^2
  [c_{\tilde\alpha}\,s_b\,s_c + s_{\tilde\alpha}\,c_c]^2
  [c_{\tilde\alpha}\,s_b\,c_c - s_{\tilde\alpha}\,s_c]^2.
\end{align}
We note that both these quantities possess the same symmetries
(\ref{Eq:xi-symm}) as $\xi_V$.
Also, $\tilde\gamma_b$ is obtained from $\tilde\gamma_t$ by the substitutions
\begin{equation}  \label{Eq:gamma_t-gamma_b}
(s_{\tilde\alpha}\leftrightarrow c_{\tilde\alpha}),\quad
(s_c\leftrightarrow c_c):\qquad \tilde\gamma_t\leftrightarrow\tilde\gamma_b.
\end{equation}

%%%%%%%%%%%%%%%%%%%%%%%%%%%%%%%%%%%%%%%%%%%%%%%%%%%%%%%%%%%%%%%%%%%%%%%%

%%%%%%%%%%%%%%%%%%%%%%%%%%%%%%%%%%%%%%%%%%%%%%%%%%%%%%%%%%%%%%%%%%%%%%%%
\begin{figure}[htb]
\refstepcounter{figure}
\label{Fig:xi-t-albc}
\addtocounter{figure}{-1}
\begin{center}
\setlength{\unitlength}{1cm}
\begin{picture}(15.0,12.0)
\put(0,-0.7)
{\mbox{\epsfysize=13cm
 \epsffile{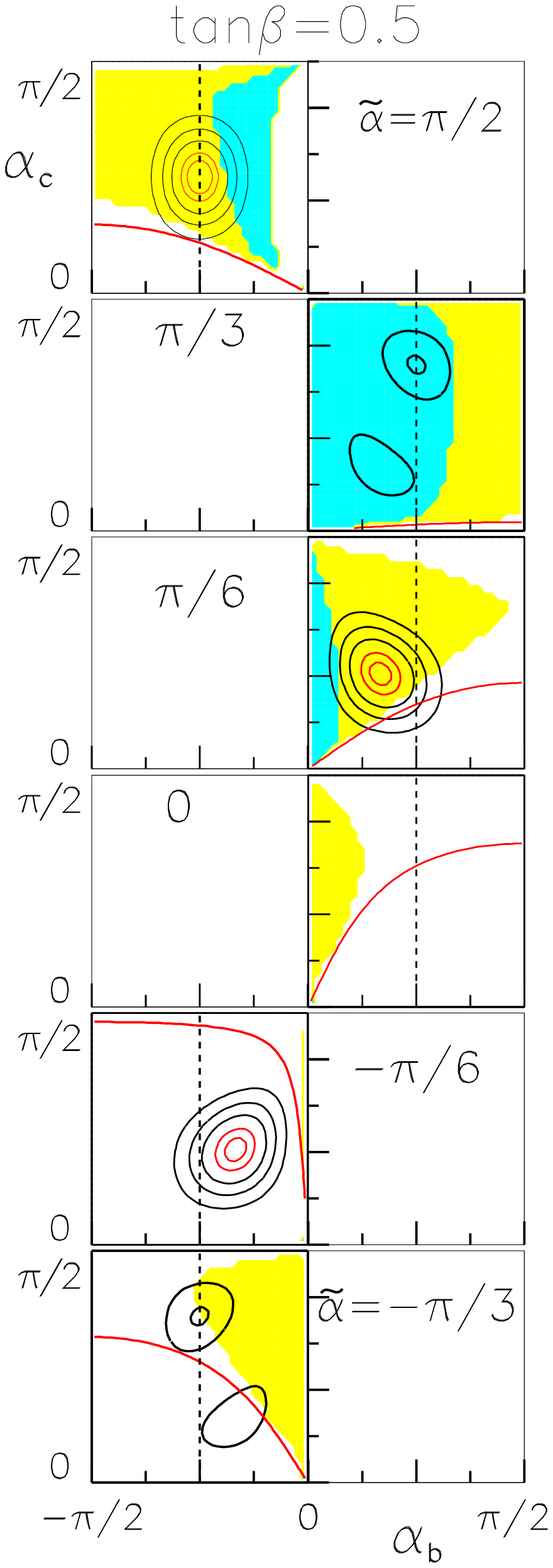}}
 \mbox{\epsfysize=13cm
 \epsffile{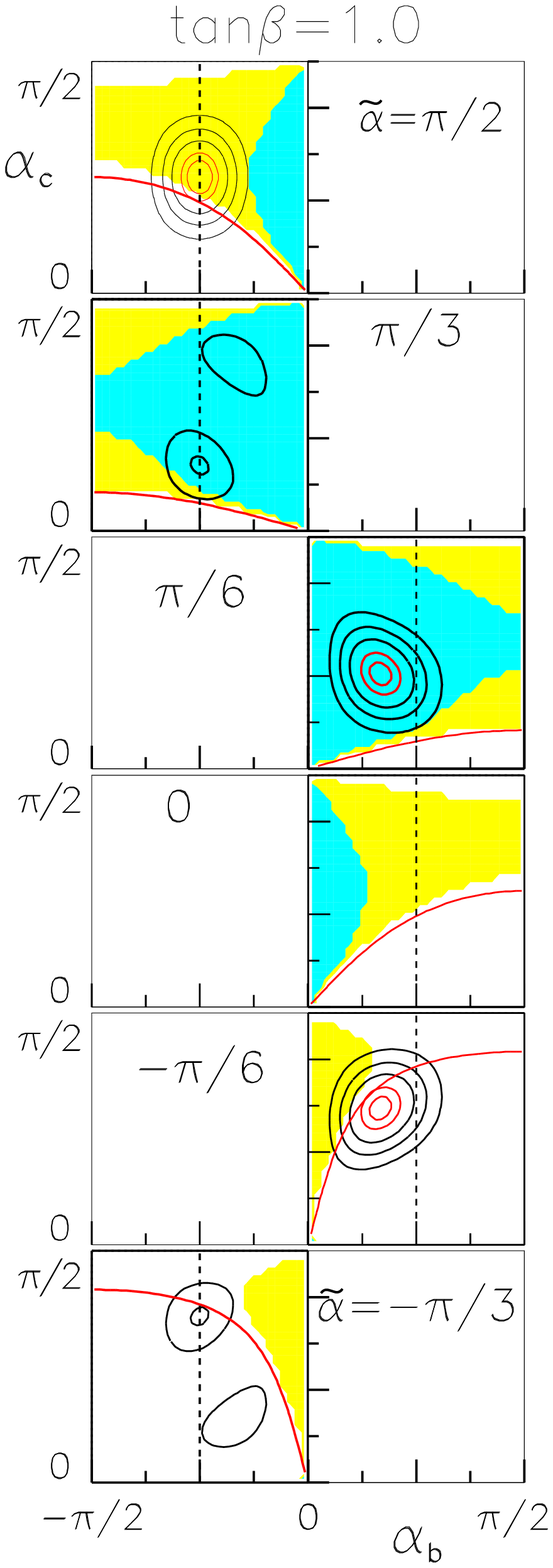}}
 \mbox{\epsfysize=13cm
 \epsffile{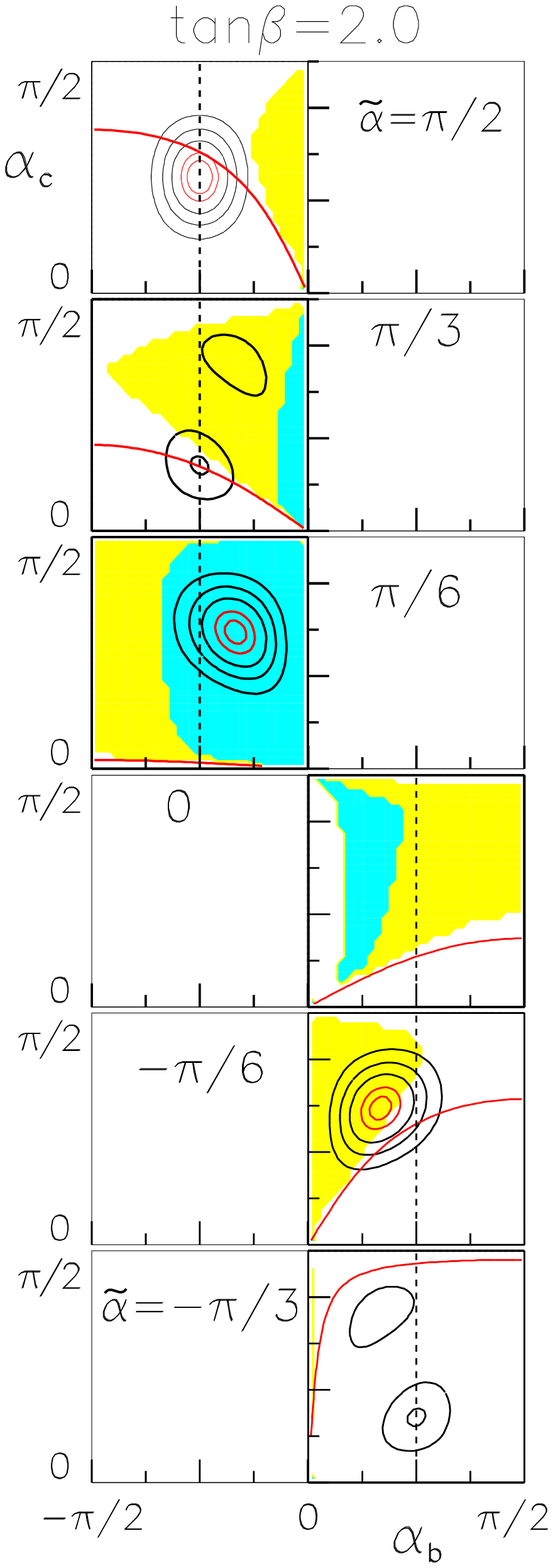}}}
\end{picture}
%\vspace*{-4mm}
\caption{Contours of constant $\gamma_t$ [see Eq.~(\ref{Eq:gamma_tb})] in
the $\alpha_b$--$\alpha_c$ plane for various values of $\tan\beta$ and
$\tilde\alpha$.  Soft parameters: $M_1=100~\text{GeV}$, $M_2=500~\text{GeV}$,
$M_{H^\pm}=600~\text{GeV}$, $\mu=300~\text{GeV}$. Dark (blue): $\xi_{\rm
pert}=1$, light (yellow): $\xi_{\rm pert}=5$.}
\end{center}
\end{figure}
%%%%%%%%%%%%%%%%%%%%%%%%%%%%%%%%%%%%%%%%%%%%%%%%%%%%%%%%%%%%%%%%%%%%%%
%%%%%%%%%%%%%%%%%%%%%%%%%%%%%%%%%%%%%%%%%%%%%%%%%%%%%%%%%%%%%%%%%%%%%%%%
\subsection{\boldmath Maxima of $\gamma_t$}
%%%%%%%%%%%%%%%%%%%%%%%%%%%%%%%%%%%%%%%%%%%%%%%%%%%%%%%%%%%%%%%%%%%%%%%%

Let us now consider the maxima of $\tilde\gamma_t$ in (\ref{Eq:Yukawa-gamma}).
We find the maximum value $\tilde\gamma_t^{\rm max}=1/1024$ for
\begin{equation}  \label{Eq-max-gam_t-caseI}
\text{Case I}:\quad \tilde\alpha=\half\pi,\quad
\alpha_b=\pm\fourth\pi,\quad
\alpha_c=\pm\fourth\pi,
\end{equation}
where the two signs are independent, and at
\begin{eqnarray} \label{Eq-max-gam_t-caseII}
\text{Case II}:\quad 
\tilde\alpha=\pm\arctan\frac{1}{\sqrt2}\ (\tilde\alpha=\pm0.196\,\pi), \quad
\alpha_b=\pm\sixth\pi, \quad \text{with} \nonumber \\
\alpha_c=\pm\arctan\frac{1}{\sqrt2}\ (\alpha_c=\pm0.196\,\pi)\quad 
\text{or}\quad
\alpha_c=\mp\arctan\sqrt2\ (\alpha_c=\pm0.304\,\pi).
\end{eqnarray}
For Case~II, the signs are subject to the constraint
$\tilde\alpha\alpha_b\alpha_c>0$ for the first $\alpha_c$ solution, and
$\tilde\alpha\alpha_b\alpha_c<0$ for the second $\alpha_c$ solution.  The
maxima of $\tilde\gamma_b$ are obtained by the substitutions
(\ref{Eq:gamma_t-gamma_b}).

Thus, it is natural to define normalized quantities
\begin{equation} \label{Eq:gamma_tb}
\gamma_t=1024\prod_{i=1}^3[R_{i2}\,R_{i3}]^2, \quad
\gamma_b=1024\prod_{i=1}^3[R_{i1}\,R_{i3}]^2,
\end{equation}
satisfying
\begin{equation}
0\le\gamma_t\le1, \quad 0\le\gamma_b\le1,
\end{equation}
as measures of CP nonconservation in the up- and down-quark sectors,
respectively. Contours of constant $\gamma_t$ are shown in the
$\alpha_b$--$\alpha_c$-plane in Fig.~\ref{Fig:xi-t-albc}.

Let us now keep $\tilde\alpha$ fixed.
Then, the maxima of $\gamma_t$ are at
\begin{alignat}{3}
&\text{Case I}:\quad \alpha_b&=\epsilon_b\,\fourth\pi, &\quad
\alpha_c=\epsilon_c\arctan\Bigl[
\sqrt{2}\Bigl(\sqrt{\tan^{-2}\tilde\alpha+\half}
+\epsilon_b\,\epsilon_c\,\tan^{-1}\tilde\alpha\Bigr)\Bigr], \nonumber \\
&\text{Case II}:\quad \alpha_b&=\epsilon_b\,\sixth\pi, &\quad
\alpha_c=\epsilon_c\arctan\Bigl[\half\Bigl(\sqrt{\tan^2\tilde\alpha+4}
-\epsilon_b\,\epsilon_c\,\tan\tilde\alpha\Bigr)\Bigr],
\end{alignat}
where $\epsilon_b$ and $\epsilon_c$ are independent sign factors:
$\epsilon_b=\pm1$, $\epsilon_c=\pm1$.
For Case~I, the corresponding maximum is (same for all sign choices)
\begin{equation}
\gamma_t=\sin^6\tilde\alpha,
\end{equation}
in agreement with Eq.~(\ref{Eq-max-gam_t-caseI}),
whereas for Case~II, the corresponding maximum is (same for all sign choices)
\begin{equation}
\gamma_t=\frac{27}{4}\,\frac{\tan^2\tilde\alpha}{(1+\tan^2\tilde\alpha)^3},
\end{equation}
which becomes 1 for $\tan\tilde\alpha=\pm1/\sqrt{2}$,
in agreement with Eq.~(\ref{Eq-max-gam_t-caseII}).
%%%%%%%%%%%%%%%%%%%%%%%%%%%%%%%%%%%%%%%%%%%%%%%%%%%%%%%%%%%%%%%%%%%%%%%%
\subsection{\boldmath Maxima of $\xi_Y$}
%%%%%%%%%%%%%%%%%%%%%%%%%%%%%%%%%%%%%%%%%%%%%%%%%%%%%%%%%%%%%%%%%%%%%%%%

While $\xi_t$ and $\xi_b$ individually diverge
as $\beta\to0$ and  $\beta\to\pi/2$, respectively,
the product over couplings to up-type and down-type
quarks is less divergent.
We define, analogous to (\ref{Eq:xi-v-def})
and (\ref{Eq:xi_t-def})
\begin{equation}  \label{Eq:xi_Y-def}
\xi_Y\equiv\xi_t\,\xi_b
\equiv\frac{1}{(\cos\beta\sin\beta)^6}\,\gamma_Y,
\end{equation}
with
\begin{equation}  \label{Eq:gamma_Y-def}
\gamma_Y=\gamma_0\,\tilde\gamma_t\,\tilde\gamma_b
=\gamma_0\,\prod_{i=1}^3\left[R_{i1}\,R_{i2}\,R_{i3}^2\right]^2.
\end{equation}
satisfying
\begin{equation}
0\le\gamma_Y\le1.
\end{equation}

%%%%%%%%%%%%%%%%%%%%%%%%%%%%%%%%%%%%%%%%%%%%%%%%%%%%%%%%%%%%%%%%%%%%%%%%
\begin{figure}[htb]
\refstepcounter{figure}
\label{Fig:xi-a-albc}
\addtocounter{figure}{-1}
\begin{center}
\setlength{\unitlength}{1cm}
\begin{picture}(15.0,12.0)
\put(0,-0.7)
{\mbox{\epsfysize=13cm
 \epsffile{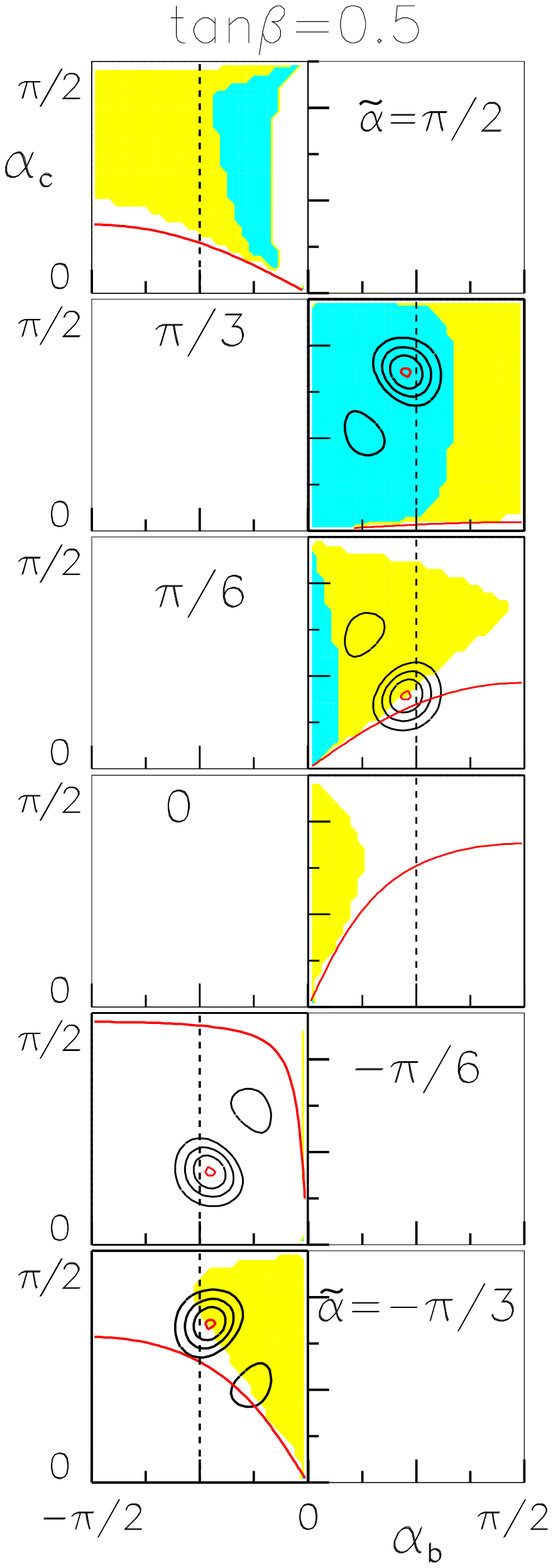}}
 \mbox{\epsfysize=13cm
 \epsffile{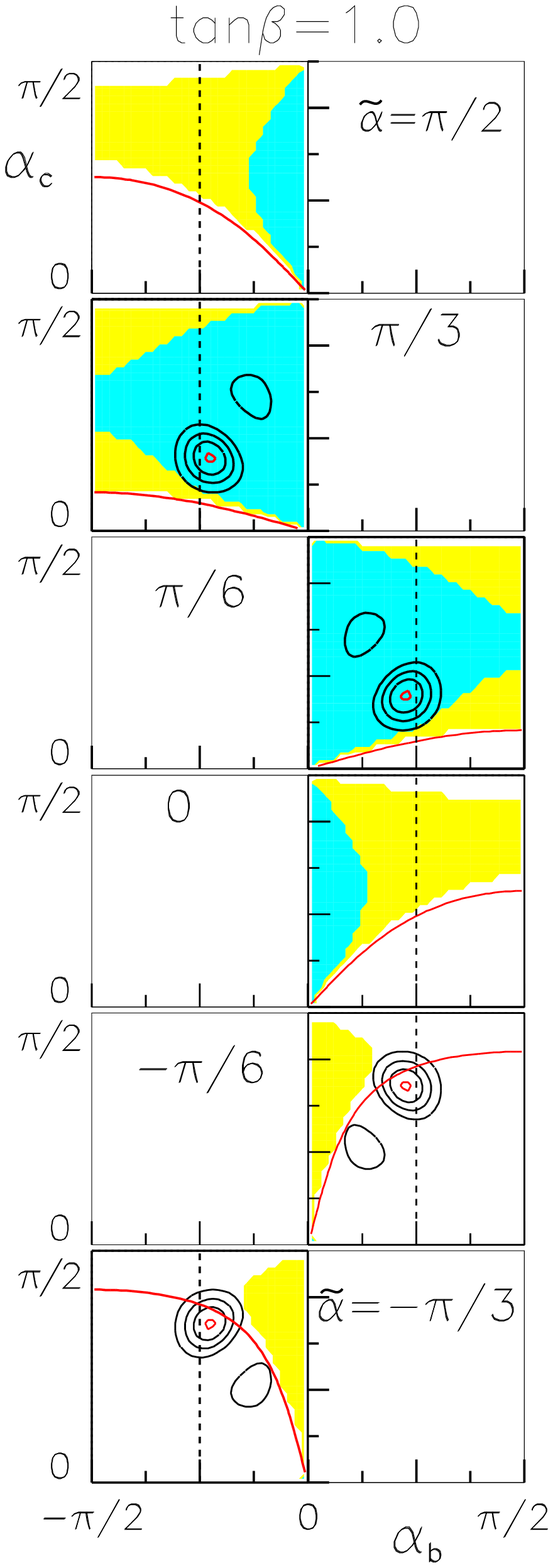}}
 \mbox{\epsfysize=13cm
 \epsffile{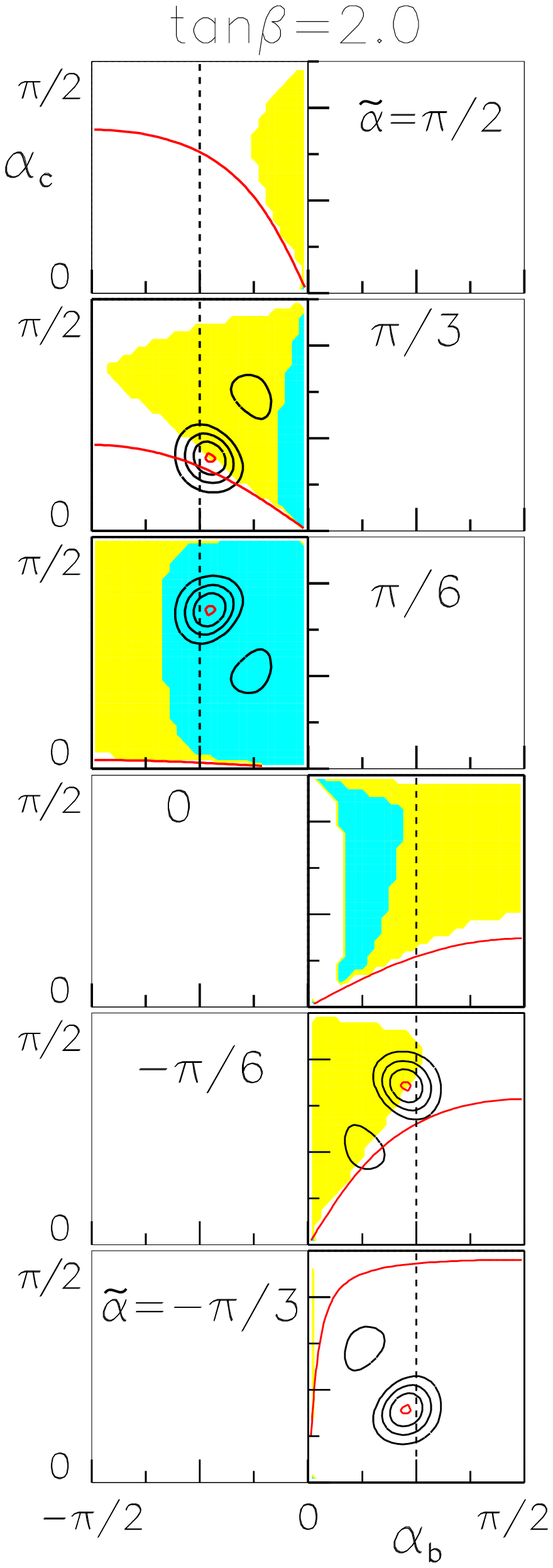}}}
\end{picture}
%\vspace*{-4mm}
\caption{Contours of constant $\gamma_Y$ in the $\alpha_b$--$\alpha_c$ plane
for various values of $\tan\beta$ and $\tilde\alpha$.  Soft parameters:
$M_1=100~\text{GeV}$, $M_2=500~\text{GeV}$, $M_{H^\pm}=600~\text{GeV}$,
$\mu=300~\text{GeV}$. Dark (blue): $\xi_{\rm pert}=1$, light (yellow):
$\xi_{\rm pert}=5$.}
\end{center}
\end{figure}
%%%%%%%%%%%%%%%%%%%%%%%%%%%%%%%%%%%%%%%%%%%%%%%%%%%%%%%%%%%%%%%%%%%%%%

Substituting from (\ref{Eq:Yukawa-gamma}), we obtain
\begin{align}  \label{Eq:xi_Y}
\gamma_Y&=\gamma_0\,c_b^{12}(c_c\,s_c\,s_b)^4
(c_{\tilde\alpha}\,s_{\tilde\alpha})^2
[s_{\tilde\alpha}\,c_c\,s_b + c_{\tilde\alpha}\,s_c]^2
[c_{\tilde\alpha}\,c_c\,s_b - s_{\tilde\alpha}\,s_c]^2
\nonumber \\
&\times[c_{\tilde\alpha}\,s_b\,s_c + s_{\tilde\alpha}\,c_c]^2
       [s_{\tilde\alpha}\,s_b\,s_c - c_{\tilde\alpha}\,c_c]^2.
\end{align}
This has a maximum for (see Appendix~A)
\begin{equation}
\tilde\alpha=\pm\fourth\pi,\qquad
\alpha_b=\pm\arcsin\sqrt{\sixth}=\pm0.13386\pi\quad (24.1\,{}^\circ),\qquad
\alpha_c=\pm\fourth\pi,
\end{equation}
with
\begin{equation} \label{Eq:gamma_0-value}
\gamma_0=\frac{2^{26}\,3^{12}}{5^{10}}
=\left(\frac{8\times 1024\times27^{2}}{3125}\right)^2
=3.652\times10^6.
\end{equation}

Fig.~\ref{Fig:xi-a-albc} exhibits contours of constant $\gamma_Y$ for some
values of $\tilde\alpha$ other than that of the maximum,
$\tilde\alpha=\fourth\pi$, in relation to the physically allowed (dark,
shaded) regions in the $\alpha_b$--$\alpha_c$ plane.  Note that $\gamma_Y$
vanishes when $\tilde\alpha=0$ or $\tilde\alpha=\pm\pi/2$, as well as on the
edges of the quadrants: $\alpha_b=0$ or $\pm\pi/2$, $\alpha_c=0$ or
$\pm\pi/2$.  Also, we note that there are secondary, local, maxima.

Although $\gamma_Y$ is, by definition, independent of $\beta$,
Fig.~\ref{Fig:xi-a-albc} shows contours of constant $\gamma_Y$ superimposed on
allowed regions for different values of $\tan\beta$, since the `shapes' and
locations of the physically allowed regions in the $\alpha_b$--$\alpha_c$
plane depend on $\tan\beta$. Accordingly, the positions of the
maxima\footnote{This is not a `maximum' in the same sense as above, since
$\tilde\alpha$ is held fixed.} of $\gamma_Y$, w.r.t. the physically allowed
regions in the $\alpha_b$--$\alpha_c$ plane are different for different values
of $\tan\beta$. For example, consider $\tilde\alpha=\pi/6$. We see from
Fig.~\ref{Fig:xi-a-albc} that for $\tan\beta=0.5$, $\gamma_Y^{\text{max}}$ is
located {\it outside} the physically allowed region while for $\tan\beta=1.0$,
this is not the case. Moreover, for $\tan\beta=2.0$, the physically allowed
region shifts the location to the `other' quadrant. To sum up, for
$\tilde\alpha=\pi/6$, the location of $\gamma_Y^{\text{max}}$ occurs at
\begin{center}
$(\alpha_b, \alpha_c)|_{\tan\beta=0.5}=(\alpha_b,
\alpha_c)|_{\tan\beta=1.0}=(-\alpha_b,
\pi/2-\alpha_c)|_{\tan\beta=2.0}$. 
\end{center}
%%%%%%%%%%%%%%%%%%%%%%%%%%%%%%%%%%%%%%%%%%%%%%%%%%%%%%%%%%%%%%%%%%%%%%%%
\subsection{\boldmath Maximizing $\zeta_t$} \label{subsect:zeta}
%%%%%%%%%%%%%%%%%%%%%%%%%%%%%%%%%%%%%%%%%%%%%%%%%%%%%%%%%%%%%%%%%%%%%%%%
We now return to the quantity $\zeta_t$ of Eq.~(\ref{Eq:zeta-def}),
which we rewrite as
\begin{equation}
\zeta_t=\left(\cos\beta/\sin^2\beta\right)^2\tilde\zeta_t
\end{equation}
with
\begin{equation}
\tilde\zeta_t=2\sum_{i=1}^3[R_{i2}\,R_{i3}]^2,  
\quad 0<\tilde\zeta_t<1.
\end{equation}

Substituting from (\ref{Eq:R-angles}) and utilising trigonometric
identities, we find  
\begin{align} \label{Eq:Yukawa-zeta}
\tilde\zeta_t&=\fourth c_b^{2}[(1-c_{2\tilde\alpha})(7+c_{4c})s_b^2 
                              +2s_{2\tilde\alpha}s_{4c}s_b
                              +(1+c_{2\tilde\alpha})(1-c_{4c})].
\end{align}

To maximize $\tilde\zeta_t$, we differentiate w.r.t.\ $\tilde\alpha$,
$\alpha_b$ and $\alpha_c$ and get:
\begin{align} \label{Eq:max-zeta}
s_{2\tilde\alpha}(7+c_{4c})s_b^2 + 2c_{2\tilde\alpha}s_{4c}s_b
-s_{2\tilde\alpha}(1-c_{4c})&=0, \nonumber \\
% The b-part
2(1-c_{2\tilde\alpha})(7+c_{4c})s_b^3 
+3s_{2\tilde\alpha}s_{4c}s_b^2 
- 2(3-4c_{2\tilde\alpha}+c_{4c})s_b 
-s_{2\tilde\alpha}s_{4c}&=0, \nonumber \\
% The c-part
(1-c_{2\tilde\alpha})s_{4c}s_b^2 - 2s_{2\tilde\alpha}c_{4c}s_b
-(1+c_{2\tilde\alpha})s_{4c} &=0.
\end{align}
Solving the three equations, one finds: $\tilde\zeta_t=1$ for
\begin{align} \label{Eq:max-zeta_alphas}
\text{Case I}:\quad 
c_{2\tilde\alpha}&=1, \quad s_b=0, \quad c_{4c}=-1 \nonumber \\
\text{Case II}:\quad 
c_{2\tilde\alpha}&=-1, \quad s_b=\pm1/\sqrt{2}, \quad c_{4c}=1
\end{align}
with the corresponding angles
\begin{align}
\text{Case I}:\quad 
\tilde\alpha=0, \quad
\alpha_b=0 \text{ or } \alpha_b=\pm\pi,\quad
\alpha_c=\pm \fourth\pi, \nonumber \\
\text{Case II}:\quad \label{Eq:max-zeta_alphas2}
\tilde\alpha=\pm\half\pi, \quad
\alpha_b=\pm\fourth\pi, \quad
\alpha_c=0\text{ or }\pm \half\pi.
\end{align}
%%%%%%%%%%%%%%%%%%%%%%%%%%%%%%%%%%%%%%%%%%%%%%%%%%%%%%%%%%%%%%%%%%%%%%%%
\begin{figure}[htb]
\refstepcounter{figure}
\label{Fig:zt-t-albc}
\addtocounter{figure}{-1}
\begin{center}
\setlength{\unitlength}{1cm}
\begin{picture}(15.0,12.0)
\put(0,-0.7)
{\mbox{\epsfysize=13cm
 \epsffile{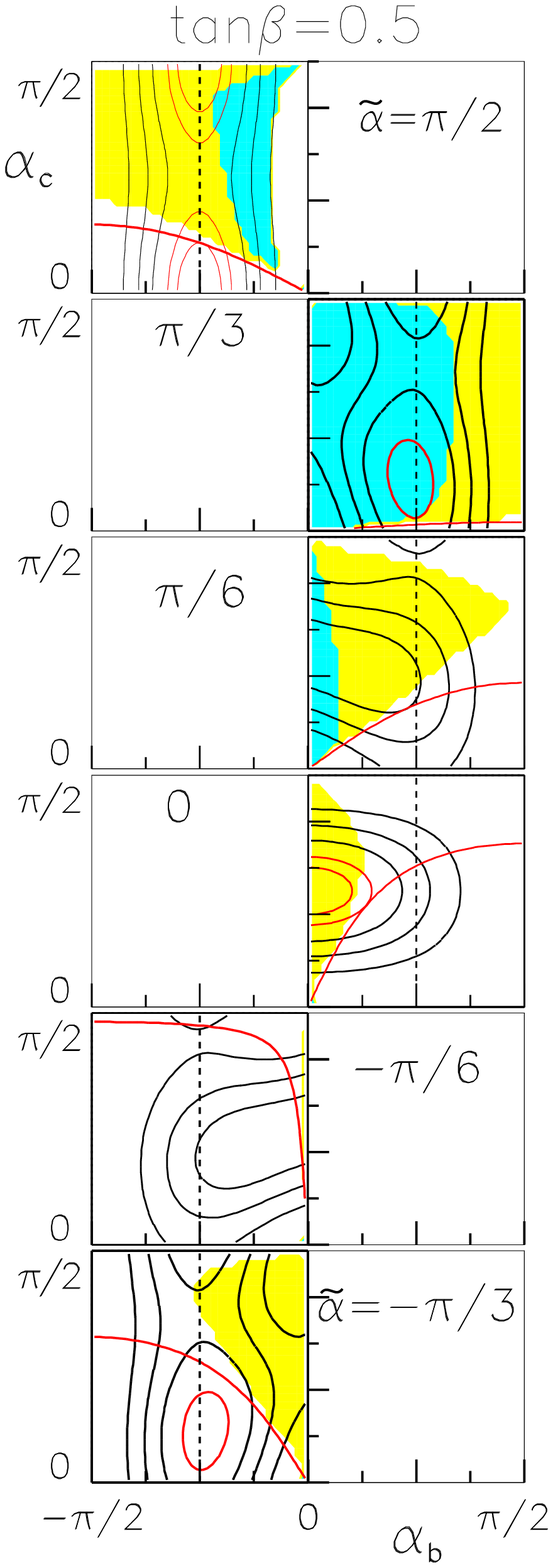}}
 \mbox{\epsfysize=13cm
 \epsffile{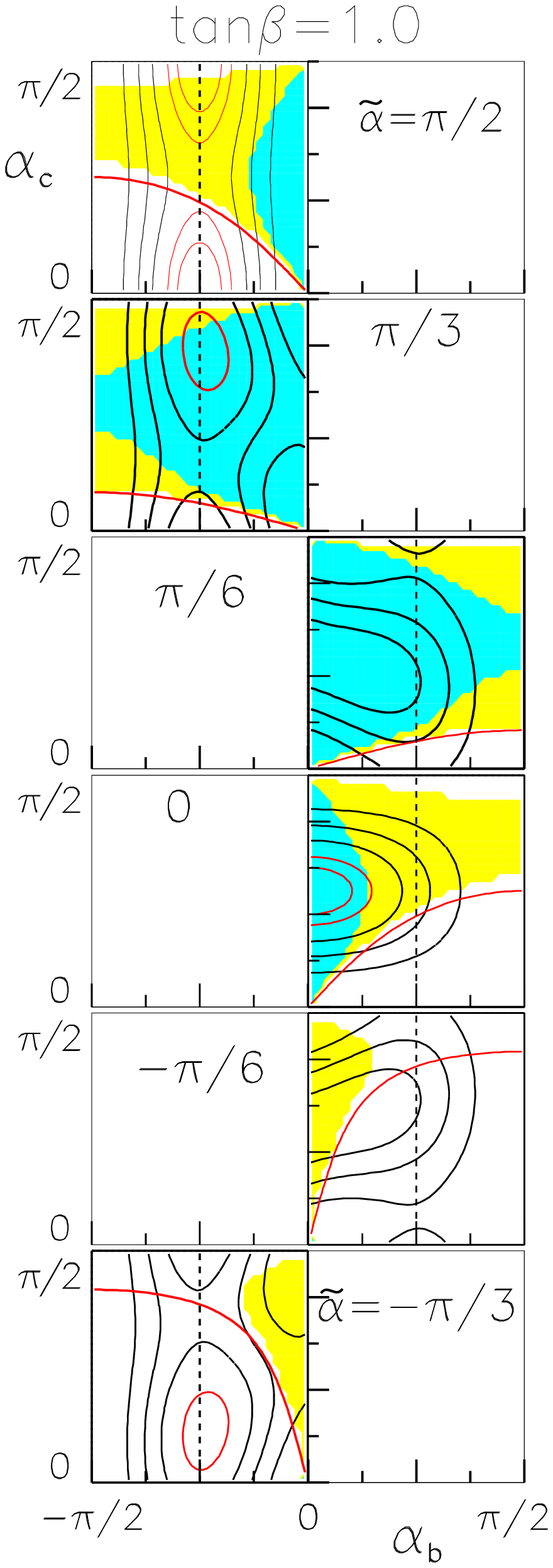}}
 \mbox{\epsfysize=13cm
 \epsffile{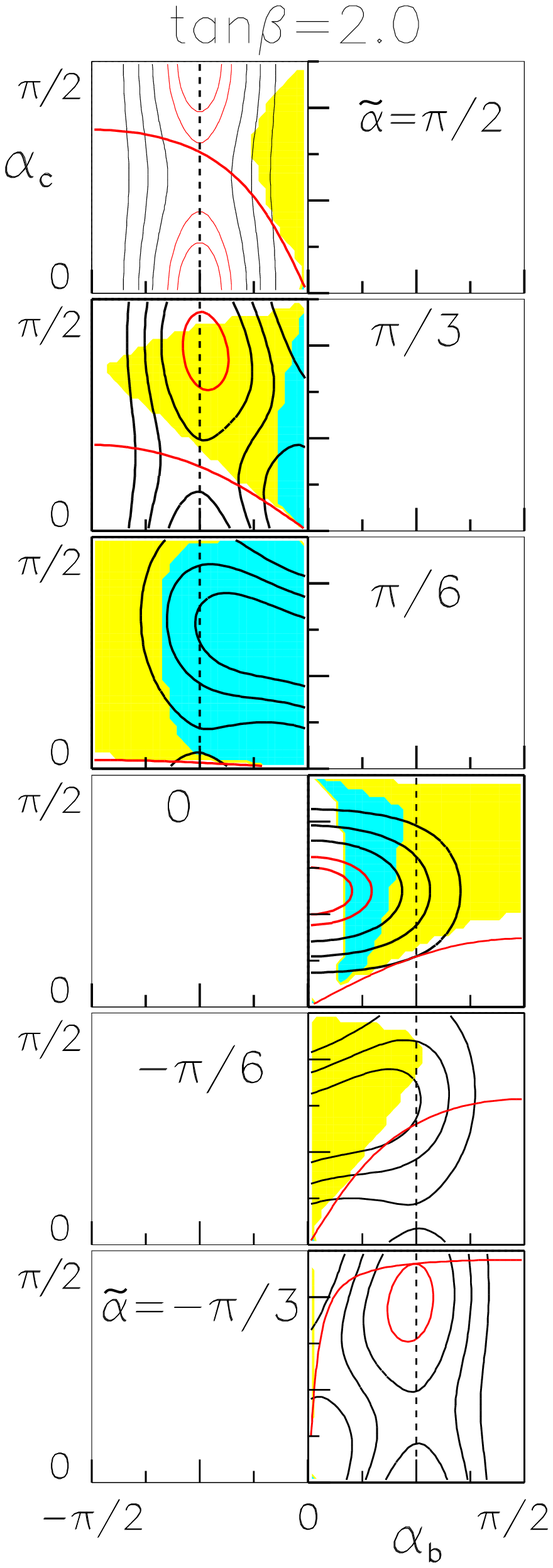}}}
\end{picture}
%\vspace*{-4mm}
\caption{Contours of constant $\tilde\zeta_t$ [see Eq.~(\ref{Eq:Yukawa-zeta})]
in the $\alpha_b$--$\alpha_c$ plane for various values of $\tan\beta$ and
$\tilde\alpha$.  Soft parameters: $M_1=100~\text{GeV}$, $M_2=500~\text{GeV}$,
$M_{H^\pm}=600~\text{GeV}$, $\mu=300~\text{GeV}$. Dark (blue): $\xi_{\rm
pert}=1$, light (yellow): $\xi_{\rm pert}=5$.}
\end{center}
\end{figure}
%%%%%%%%%%%%%%%%%%%%%%%%%%%%%%%%%%%%%%%%%%%%%%%%%%%%%%%%%%%%%%%%%%%%%%

Considering now $\tilde\alpha$ fixed, we find the maxima:
\begin{align}
\text{Case I}:\quad
\alpha_b=0\  (\text{or }\pm\pi), \quad
\alpha_c=\pm\fourth\pi, 
\end{align}
for which 
\begin{align}
\tilde\zeta_t=\half(1+c_{2\tilde\alpha})
\end{align}
coincides with Case I in (\ref{Eq:max-zeta_alphas}) for 
$c_{2\tilde\alpha}=1$, and
\begin{align}
\text{Case II}:\quad
\alpha_b=\pm\arcsin
\frac{\sqrt{3 - 5c_{2\tilde\alpha}}}{2\sqrt{2}\sqrt{1 - c_{2\tilde\alpha}}}, 
\quad
\alpha_c=\pm\frac{1}{4}\biggl[\pi-\arccos
\frac{5 + 13c_{2\tilde\alpha}}{11+3c_{2\tilde\alpha}}\biggr],
\end{align}
provided $c_{2\tilde\alpha}\le3/5$.
In this case
\begin{align}
\tilde\zeta_t&=
\frac{(5 - 3c_{2\tilde\alpha})}
{32(1 - c_{2\tilde\alpha})(11+3c_{2\tilde\alpha})^2}
\bigl[(11+3c_{2\tilde\alpha})
(43-10c_{2\tilde\alpha}+11c_{2\tilde\alpha}^2)\nonumber \\
&+(44+12c_{2\tilde\alpha})\sqrt{3 - 5c_{2\tilde\alpha}}
\sqrt{1+c_{2\tilde\alpha}}
\sqrt{3-2c_{2\tilde\alpha}-5c_{2\tilde\alpha}^2}\bigr],
\end{align}
agrees with Case II in (\ref{Eq:max-zeta_alphas}) for $c_{2\tilde\alpha}=-1$.

Fig.~\ref{Fig:zt-t-albc} exhibits contours of constant $\tilde\zeta_t$ in the
$\alpha_b$--$\alpha_c$ plane for selected values of $\tan\beta$ and
$\tilde\alpha$.  We read off from Fig.~\ref{Fig:zt-t-albc} that for
$\tilde\alpha=0$, the quantity $\tilde\zeta_t$ takes its maximum value at
$(\alpha_b, \alpha_c)=(0, \fourth\pi)$ which again is consistent with Case~I
in (\ref{Eq:max-zeta_alphas}). For particular values of $\tilde\alpha$ and
$\alpha_c$, there are also saddle points, for example at $(\tilde\alpha,
\alpha_b, \alpha_c)=(\half\pi, -\fourth\pi, \fourth\pi)$.  For a given value
of $\tilde\alpha$, these saddle points are located at
\begin{align}
\alpha_b=\pm\fourth\pi,\quad
\alpha_c=\pm\fourth\arccos
\left(\frac{1+3c_{2\tilde\alpha}}{3+c_{2\tilde\alpha}}\right)
\end{align}
in the $\alpha_b$--$\alpha_c$ plane.

In the top-Higgs Yukawa sector, $\gamma_t$ [see Eq.~(\ref{Eq:gamma_tb})] and
$\tilde\zeta_t$ are both sizable for large $\tilde\alpha$ and
$|\alpha_b|\simeq\pi/4$, as we see in Figs.~\ref{Fig:xi-t-albc} and
\ref{Fig:zt-t-albc}.  However, the two measures have different features.  For
example, for the same value of $\tilde\alpha=0$, where $\tilde\zeta_t$ has a
maximum, $\gamma_t$ vanishes. Moreover, for $\alpha_b=0$, $\gamma_t$ vanishes
(regardless the values of $\tilde\alpha$ and $\alpha_c$) while $\tilde\zeta_t$
takes its maximum value (for $\tilde\alpha=0$ and $\alpha_c=\fourth\pi$). This
again shows that these quantities $\gamma_t$ and $\tilde\zeta_t$ behave rather
differently for a given set of the angles $(\tilde\alpha, \alpha_b,
\alpha_c)$.

%%%%%%%%%%%%%%%%%%%%%%%%%%%%%%%%%%%%%%%%%%%%%%%%%%%%%%%%%%%%%%%%%%%%%%%%
\section{\boldmath CP nonconservation in $pp\to t\bar t$}
\setcounter{equation}{0}
%%%%%%%%%%%%%%%%%%%%%%%%%%%%%%%%%%%%%%%%%%%%%%%%%%%%%%%%%%%%%%%%%%%%%%%%
The above studies refer to the tree-level couplings of Higgs particles to
vector particles and fermions. These are difficult to study directly, since
the Higgs particles as well as the vector particles and the relevant fermions
are unstable. The implication is that it is easier to access these couplings
via various loop effects.  We shall here consider one such example, namely the
production amplitudes for the $t\bar t$ through
gluon fusion, where CP nonconservation is induced by 
non-standard neutral Higgs exchange.

CP nonconservation in the production of $t\bar t$ pairs at future hadronic
colliders has been studied in considerable detail \cite{Bernreuther:1994hq}.
For a detailed application to the 2HDM, see also \cite{Khater:2003wq}.

One process of particular interest is
\begin{equation}
pp\to t\bar t X,
\end{equation}
where the $t$ and $\bar t$ decay semileptonically, and the lepton energy
difference is measured \cite{Bernreuther:1994hq,Khater:2003wq}:
\begin{equation}  \label{Eq:A1}
A_1=E_+-E_-.
\end{equation}
(For a discussion of other observables, see
\cite{Bernreuther:1994hq,Bernreuther:1998qv}.)  The expectation value of this
observable will in general be non-zero if there between the quarks in the
final state are exchanges of Higgs bosons that are not eigenstates under CP.
The quantity [see Eq.~(\ref{Eq:H_itt})]
\begin{equation}   \label{Eq:gamma_CP-i}
\gamma_{CP,j}
=-a_j^{(t)} \tilde a_j^{(t)}
=\frac{\cos\beta}{\sin^2\beta}\, R_{j2}R_{j3}
\end{equation}
then plays a crucial role, together with non-trivial functions of the
kinematics (given by the loop integrals). 

If the neutral-Higgs spectrum has a large gap between the lightest Higgs boson
and the next one, then the lightest one will give the dominant
contribution to $A_1$, and the
amount of CP nonconservation is roughly proportional to
\begin{equation}   \label{Eq:gamma_CP-1}
\gamma_{CP,1}
=\half\,\frac{\sin\tilde\alpha\sin(2\alpha_b)}{\tan\beta\sin\beta},
\end{equation}
which is maximized for small $\tan\beta$ and for $(\tilde\alpha, \alpha_b)=
(\pm\pi/2, \pm\pi/4)$, corresponding to the dashed lines at
$\alpha_b=\pm\pi/4$ in Figs.~\ref{Fig:albc-100-500-600-300},
\ref{Fig:xi-V-albc}, \ref{Fig:xi-t-albc}, \ref{Fig:xi-a-albc} and
\ref{Fig:zt-t-albc}. These values as well, $(\tilde\alpha, \alpha_b)=
(\pm\pi/2, \pm\pi/4)$, together with $\alpha_c=0\text{ or }\pm \half\pi$,
coincide with those of Case~II [see Eq.~(\ref{Eq:max-zeta_alphas2})] that
maximize $\tilde\zeta_t$. Furthermore, $(\tilde\alpha, \alpha_b)= (\pi/2,
\pm\pi/4)$, together with $\alpha_c=\pm\fourth\pi$, coincide with those of
Case~I [see Eq.~(\ref{Eq-max-gam_t-caseI})] that maximize $\gamma_t$.  These
results indicate that large $\tilde\alpha$ together with
$|\alpha_b|\simeq\pi/4$ favour large CP nonconservation in the Yukawa sector.
It is immediately obvious that this is not compatible with the condition of
maximal CP nonconservation in the gauge--Higgs sector \cite{Mendez:1991gp},
$\xi_V=1$ [see Eqs.~(\ref{Eq:max-alb=0}) and (\ref{Eq:max-alc=0})].

In addition to the contribution from the lightest Higgs boson,
there will in general also be non-negligible contributions from the others.
Because of the orthogonality of the rotation matrix $R$, not
all $\gamma_{CP,j}$ can have the same sign, so there will be cancellations.

Let us define the `signal-to-noise ratio', or sensitivity
\cite{Bernreuther:1994hq}
\begin{equation}  \label{Eq:S/N}
\frac{S}{N}=\frac{\langle A_1\rangle}
{\sqrt{\langle A_1^2\rangle-\langle A_1\rangle^2}},
\end{equation}
which provides a measure of how much data would be required to see an effect.

%%%%%%%%%%%%%%%%%%%%%%%%%%%%%%%%%%%%%%%%%%%%%%%%%%%%%%%%%%%%%%%%%%%%%%%%
\begin{figure}[htb]
\refstepcounter{figure}
\label{Fig:a1-s}
\addtocounter{figure}{-1}
\begin{center}
\setlength{\unitlength}{1cm}
\begin{picture}(10.0,7.0)
\put(-3.5,0){
\mbox{\epsfysize=7cm
\epsffile{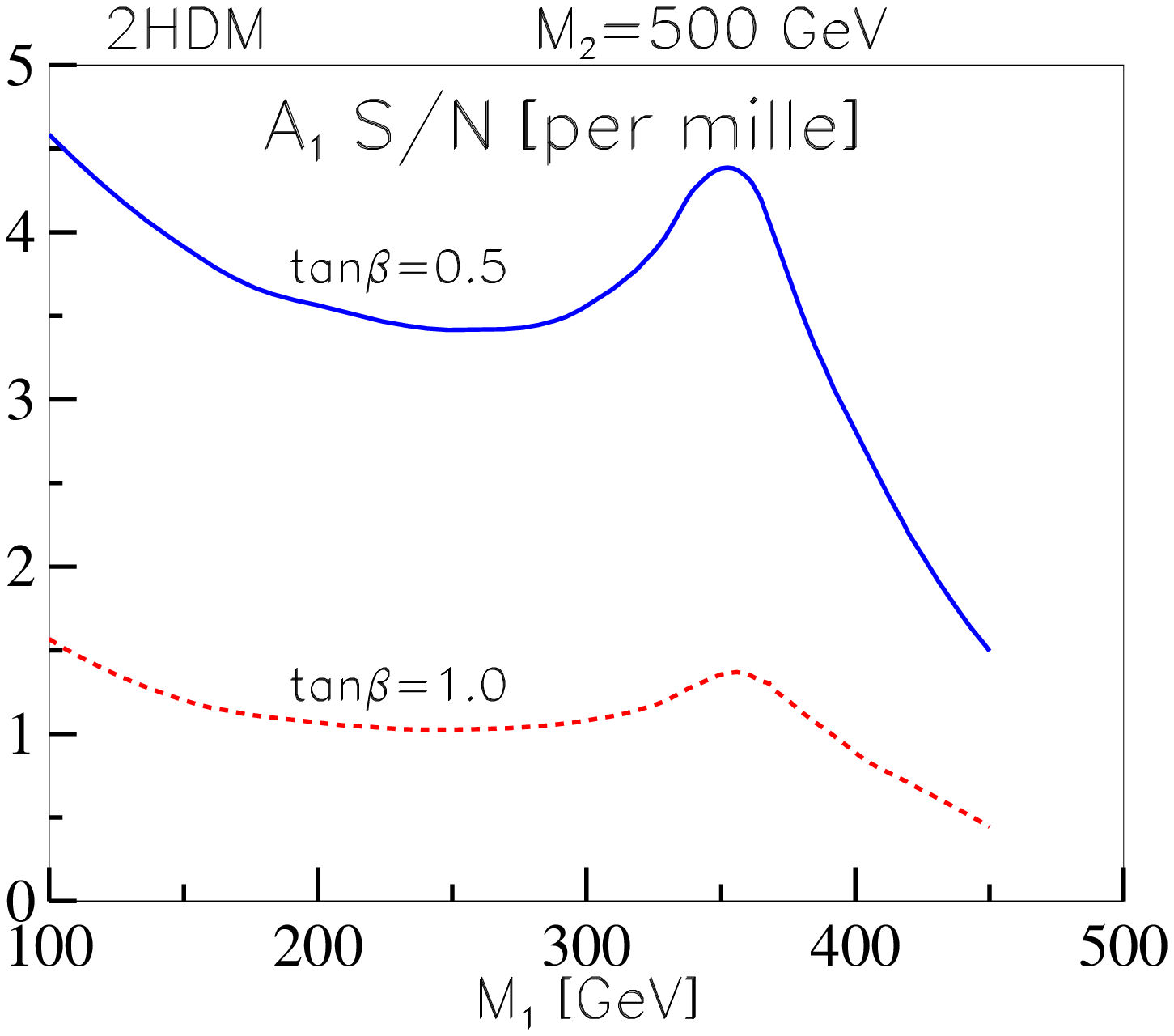}}
\mbox{\epsfysize=7cm
\epsffile{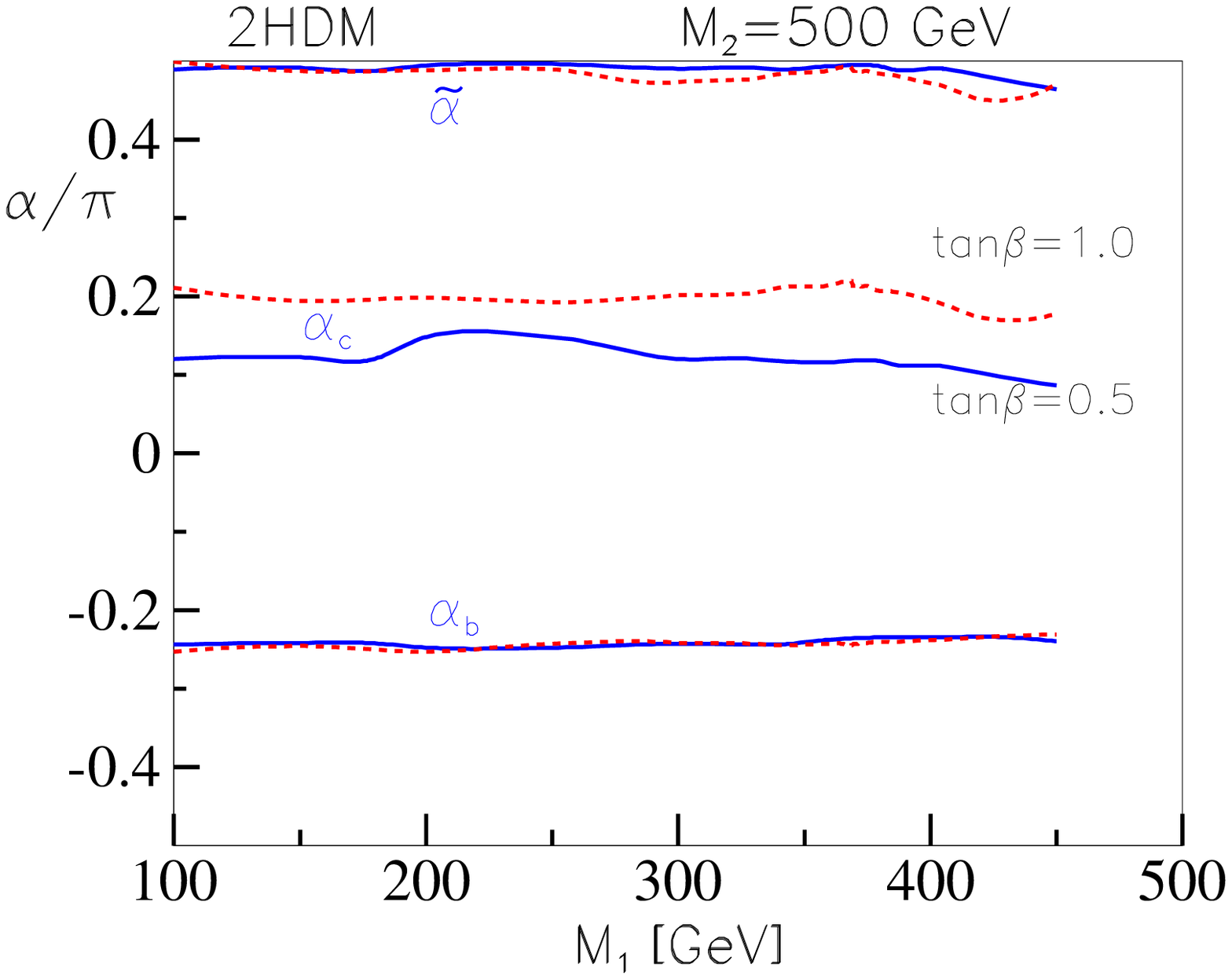}}}
\end{picture}
\vspace*{-4mm}
\caption{Left panel: Maximal sensitivity [see (\ref{Eq:S/N})]
for the observable (\ref{Eq:A1}), for fixed $M_1$, $M_2$
and two values of $\tan\beta$.
Right panel: Corresponding values of the angles $\tilde\alpha$,
$\alpha_b$ and $\alpha_c$.
Soft parameters: $M_2=500~\text{GeV}$, 
$M_{H^\pm}=600~\text{GeV}$, $\xi_{\rm pert}=5$.}
\end{center}
\end{figure}
%%%%%%%%%%%%%%%%%%%%%%%%%%%%%%%%%%%%%%%%%%%%%%%%%%%%%%%%%%%%%%%%%%%%%%

It is interesting to maximize the amount of CP nonconservation that results
for the observable $A_1$, over the relevant parameters of the model.  In
Fig.~\ref{Fig:a1-s} we show the result of such a maximization of the
sensitivity (\ref{Eq:S/N}).  The quantity $A_1$ and its spread $A_1^2$ are
computed as given in \cite{Bernreuther:1994hq,Khater:2003wq}, using the
`LoopTools' package \cite{Hahn:1998yk,vanOldenborgh:1989wn}, and convoluted
with the CTEQ6 parton distribution functions \cite{Pumplin:2002vw} for the LHC
energy of 14~TeV. The resulting quantity is then maximized using the
`MINUIT' package \cite{James:1975dr}.

The actual maximization is rather CPU-intensive:
In order to evaluate $A_1$ and $S/N$, three-dimensional integrals
(a convolution integral over the parton distribution functions,
an integral over the polar angle of the top quark
with respect to the beam,
an integral over $\hat s$, the invariant mass squared of the $t\bar t$ pair)
involving non-trivial loop functions are required. These are then 
maximized in the three angles parameterizing the
2HDM mass matrix: $\tilde\alpha$, $\alpha_b$ and $\alpha_c$
(keeping the two lowest Higgs masses fixed).

In this maximization, we have kept $M_2=500~\text{GeV}$ fixed, and considered
two values of $\tan\beta$ (0.5 and 1.0), and a range of values of $M_1$.  The
resulting angles $\tilde\alpha$ and $\alpha_b$ are rather independent of $M_1$
as well as the choice of $\tan\beta$, whereas $\alpha_c$ has some dependence
on $\tan\beta$, as shown in the right panel of Fig.~\ref{Fig:a1-s}.

For a given value of $M_1$, the resulting maximum is close to that found in
\cite{Khater:2003wq}, maximizing only with respect to the $H_1$ contribution.
We note that, considered as a function of $M_1$, 
there is a peak associated with the $t\bar t$ threshold.
This is due to the contribution of the $t\bar t$ triangle diagram
\cite{Bernreuther:1994hq,Khater:2003wq}.

As discussed in \cite{Khater:2003wq}, the heavier Higgs states have a tendency
to reduce the CP-violating effect of the lightest one, unless they are
sufficiently heavy to decouple.  Thus, for a fixed value of the lightest Higgs
mass, $M_1$, the over-all CP-nonconservation should increase as the second
Higgs boson becomes heavier.  This effect is illustrated in
Fig.~\ref{Fig:a1-s-m2} for the case of $M_1=100~\text{GeV}$ and two values of
$\tan\beta$ (0.5 and 1.0).
Apart from some wiggles due to numerical noise,
it is seen that there is a rather smooth increase of the sensitivity
as the mass gap $M_2-M_1$ increases.

%%%%%%%%%%%%%%%%%%%%%%%%%%%%%%%%%%%%%%%%%%%%%%%%%%%%%%%%%%%%%%%%%%%%%%%%
\begin{figure}[htb]
\refstepcounter{figure}
\label{Fig:a1-s-m2}
\addtocounter{figure}{-1}
\begin{center}
\setlength{\unitlength}{1cm}
\begin{picture}(10.0,7.0)
\put(1,0){
\mbox{\epsfysize=7cm
\epsffile{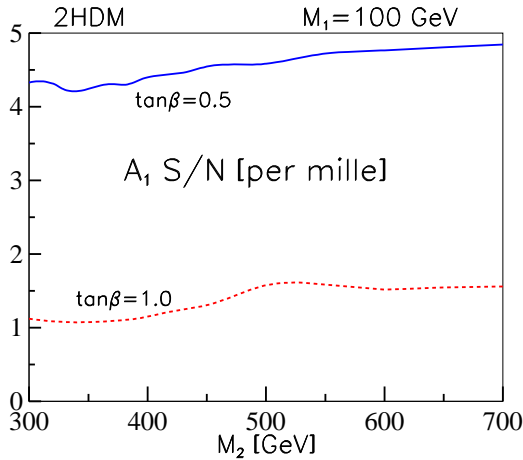}}
}
\end{picture}
\vspace*{-4mm}
\caption{Maximum sensitivity $S/N$ [see (\ref{Eq:S/N})] {\it vs.}\ $M_2$,
for fixed $M_1=100~\text{GeV}$.
Soft parameters: 
$M_{H^\pm}=700~\text{GeV}$, $\xi_{\rm pert}=5$.}
\end{center}
\end{figure}
%%%%%%%%%%%%%%%%%%%%%%%%%%%%%%%%%%%%%%%%%%%%%%%%%%%%%%%%%%%%%%%%%%%%%%

Let us now comment on the maximum CP nonconservation in the Yukawa sector, as
given by the sensitivity in the quantity $A_1$, compared with that of the
gauge-Higgs sector, $\xi_V$. We already stated that these concepts are
different. This statement can be made quantitative by considering the value of
$\xi_V$ that corresponds to the rotation angles $\tilde\alpha$, $\alpha_b$ and
$\alpha_c$ for which the sensitivity in $A_1$ is maximal.  We find that
$\xi_V\simeq0.6$ and 0.3, for $\tan\beta=0.5$ and 1.0, respectively.
%%%%%%%%%%%%%%%%%%%%%%%%%%%%%%%%%%%%%%%%%%%%%%%%%%%%%%%%%%%%%%%%%%%%%%%%
\section{Concluding remarks}
\setcounter{equation}{0}
%%%%%%%%%%%%%%%%%%%%%%%%%%%%%%%%%%%%%%%%%%%%%%%%%%%%%%%%%%%%%%%%%%%%%%%%
The concept of maximal CP nonconservation has been extended from
the gauge--Higgs sector to the Yukawa sector, where various measures for
CP nonconservation have been introduced and investigated.
Large values of $\tilde\alpha$ and $|\alpha_b|\simeq\pi/4$ favour large
CP nonconservation in the Yukawa sector. But,
in general, the maxima of CP nonconservation will in these two sectors 
not coincide. There could even be {\it maximal} CP nonconservation 
in one sector, and little or none in the other.

We have here studied the {\it simplest} version of the 2HDM that allows for 
CP nonconservation, where this CP nonconservation is given by {\it one}
parameter, namely $\Im\lambda_5$ in the potential (\ref{Eq:gko-pot}).
One could consider two more, independent parameters in the Higgs potential
that generate CP nonconservation, namely $\Im\lambda_6$ and $\Im\lambda_7$
(see, e.g., \cite{GKO}).  These terms in the potential are often considered
less attractive, since they violate the $Z_2$ symmetry of the potential by
terms which are quartic in the Higgs fields and thus make it more difficult to
control flavour-changing neutral currents \cite{Glashow:1976nt,Branco}.

However, if present, such terms would lead to a less constrained theory.
While the Yukawa couplings (for Model~II) are still given by the same elements
of the rotation matrix $R$ (and hence by the same expression in terms of
$\tan\beta$ and the rotation angles $\tilde\alpha$, $\alpha_b$ and
$\alpha_c$), the masses $M_2$ and $M_3$ would be less constrained.  By making
these masses larger, the contribution of the lightest one, $H_1$, would be a
better approximation to the over-all CP nonconservation.
\medskip

\noindent
{\bf Acknowledgments:}
It is a pleasure to thank the organizers of the Epiphany 2003 Conference for
creating a most stimulating atmosphere, and for excellent hospitality.
%%%%%%%%%%%%%%%%%%%%%%%%%%%%%%%%%%%%%%%%%%%%%%%%%%%%%%%%%%%%%%%%%%%%%%%%
\section*{Appendix A. \boldmath Maximizing $\gamma_Y$}
\renewcommand{\thesection}{A} \setcounter{equation}{0}
%%%%%%%%%%%%%%%%%%%%%%%%%%%%%%%%%%%%%%%%%%%%%%%%%%%%%%%%%%%%%%%%%%%%%%%%
This appendix deals with the maximization of $\gamma_Y$,
Eq.~(\ref{Eq:gamma_Y-def}).  We shall first rewrite
$\tilde\gamma_b\tilde\gamma_t$ in terms of double angles.  Let
\begin{align}
x&\equiv
(s_{\tilde\alpha}\,s_b\,c_c + c_{\tilde\alpha}\,s_c)
(c_{\tilde\alpha}\,s_b\,c_c - s_{\tilde\alpha}\,s_c), \nonumber\\
y&\equiv
(c_{\tilde\alpha}\,s_b\,s_c + s_{\tilde\alpha}\,c_c)
(s_{\tilde\alpha}\,s_b\,s_c - c_{\tilde\alpha}\,c_c),
\end{align}
then
\begin{equation}
\tilde\gamma_b\tilde\gamma_t=z^2
\end{equation}
with
\begin{equation}
z=c_b^6\,c_{\tilde\alpha}\, s_{\tilde\alpha}\,s_b^2(c_c s_c)^2\,xy.
\end{equation}
Maximizing $\tilde\gamma_b\tilde\gamma_t$ amounts to maximizing 
the absolute value of $z$.

We first note that
\begin{equation}
x=\fourth s_{2\tilde\alpha}[(1+s_b^2)c_{2c}-c_b^2]
+\half c_{2\tilde\alpha}\,s_b\,s_{2c}
\end{equation}
where $c_{2\tilde\alpha}=\cos(2\tilde\alpha)$, $c_{2c}=\cos(2\alpha_c)$, etc.
Furthermore, $y$ can be obtained from $x$ by the substitutions
$c_{\tilde\alpha}\leftrightarrow s_{\tilde\alpha}$ and $c_c\leftrightarrow
s_c$, implying $c_{2\tilde\alpha}\leftrightarrow-c_{2\tilde\alpha}$,
$c_{2c}\leftrightarrow-c_{2c}$, with $s_{2\tilde\alpha}$ and $s_{2c}$
unchanged.  Thus,
\begin{align}
xy&=
\{-\fourth s_{2\tilde\alpha}\,c_b^2+[\half c_{2\tilde\alpha}\,s_b\,s_{2c}
+\fourth s_{2\tilde\alpha}(1+s_b^2)c_{2c}]\} \nonumber \\
&\times
\{-\fourth s_{2\tilde\alpha}\,c_b^2 -[\half c_{2\tilde\alpha}\,s_b\,s_{2c}
+\fourth s_{2\tilde\alpha}(1+s_b^2)c_{2c}]\} \nonumber \\
&={\textstyle\frac{1}{16}}[s_{2\tilde\alpha}^2\,c_b^4
-4c_{2\tilde\alpha}^2\,s_b^2\,s_{2c}^2
-s_{2\tilde\alpha}^2(1+s_b^2)^2c_{2c}^2
-4c_{2\tilde\alpha}\,s_{2\tilde\alpha}(1+s_b^2)\,s_b\,c_{2c}\,s_{2c}] 
\end{align}

The maximum is given by the three conditions:
\begin{equation}
\frac{\partial z}{\partial\tilde\alpha}=0, \qquad
\frac{\partial z}{\partial\alpha_b}=0, \qquad
\frac{\partial z}{\partial\alpha_c}=0, \qquad
\end{equation}
or equivalently:
\begin{align}
\label{Eq:d_atil=0}
&3c_{2\tilde\alpha}\,(1-c^2_{2\tilde\alpha})(1-c_{2c}^2)(1+s_b^4)
+4s_{2\tilde\alpha}\,c_{2c}\,s_{2c}\,s_b(1-3c^2_{2\tilde\alpha})(1+s_b^2)
\nonumber \\
&+2c_{2\tilde\alpha}\,s_b^2[1-7c_{2c}^2-3c^2_{2\tilde\alpha}(1-3c_{2c}^2)]=0,
\\
\label{Eq:d_ab=0}
&(1-c^2_{2\tilde\alpha})(1-c_{2c}^2)(1-6s_b^6)
+c_{2\tilde\alpha}\,s_{2\tilde\alpha}\,c_{2c}\,s_{2c}\,s_b(22s_b^4+8s_b^2-6)
-8s_b^2(1-c^2_{2\tilde\alpha}c_{2c}^2)\nonumber\\
&+s_b^4[13-27c^2_{2\tilde\alpha}c_{2c}^2+7(c^2_{2\tilde\alpha}+c_{2c}^2)] =0,
\\
\label{Eq:d_ac=0}
&c_{2c}\,(1-c^2_{2\tilde\alpha})(1-c_{2c}^2)(1+s_b^4)
+c_{2\tilde\alpha}\,s_{2\tilde\alpha}\,s_{2c}\,s_b(1-4c_{2c}^2)(1+s_b^2)
\nonumber \\
&-2c_{2c}\,s_b^2[2c^2_{2\tilde\alpha}+c_{2c}^2(1-3c^2_{2\tilde\alpha})]=0
\end{align}

While these three equations are highly non-linear, the solution
of interest is actually obtained quite simply by setting
\begin{equation}
c_{2\tilde\alpha}=0,\quad 
c_{2c}=0,
\end{equation}
whereby Eqs.~(\ref{Eq:d_atil=0}) and (\ref{Eq:d_ac=0}) become
trivially satisfied, and Eq.~(\ref{Eq:d_ab=0}) takes the simple form
\begin{equation}
6s_b^6-13s_b^4+8s_b^2-1=0,
\end{equation}
the interesting solution of which is $s_b^2=1/6$.

Summarizing, the maxima are obtained for
\begin{equation}
\tilde\alpha=\pm\fourth\pi,\qquad
\alpha_b=\pm\arcsin\sqrt{\sixth}=\pm0.13386\pi\quad (24.1\,{}^\circ),\qquad
\alpha_c=\pm\fourth\pi,
\end{equation}
at which point
\begin{equation}
z=\pm \frac{3125}{8\times 1024\times27^{2}}
\end{equation}
determines the $\gamma_0$ of (\ref{Eq:gamma_0-value}).
%%%%%%%%%%%%%%%%%%%%%%%%%%%%%%%%%%%%%%%%%%%%%%%%%%%%%%%%%%%%%%%%%%%%%%%%

\end{document}